\documentclass{aa}  

\usepackage{float}
\usepackage{graphicx}
\usepackage{txfonts}
\usepackage{natbib}
\bibpunct{(}{)}{,}{a}{}{,}  
\usepackage{pdflscape}
\usepackage{orcidlink}
\usepackage{hyperref}
\hypersetup{breaklinks=true,
            colorlinks,
            filecolor=blue,
            linkcolor=blue,
            citecolor=blue,
            urlcolor=blue,
            anchorcolor=blue} 
\newcommand\feh{\ensuremath{[\mathrm{Fe}/\mathrm{H}]~}}
\newcommand\afeh{\ensuremath{[\alpha/\mathrm{Fe}]~}}
\newcommand\dex{\ensuremath{~\mathrm{dex}}}
\newcommand\Teff{\ensuremath{T_{\mathrm{eff}}}}
\newcommand\vt{\ensuremath{v_{\mathrm{t}}}}

\defcitealias{ceccarelli2024}{Paper I}

\begin{document}

   \title{A Walk on the Retrograde Side (WRS) project}

   \subtitle{II. Chemistry to disentangle in situ and accreted components in Thamnos}
   
   \titlerunning{A Walk on the Retrograde Side. II}

   \authorrunning{Ceccarelli, E., et al.}   
   
   \author{E. Ceccarelli
          \inst{1,2}
          \orcidlink{0009-0007-3793-9766}
          \and
          D. Massari
          \inst{1}
          \orcidlink{0000-0001-8892-4301}   
          \and
          M. Palla
          \inst{2,1}
          \orcidlink{0000-0002-3574-9578} 
          \and
          A. Mucciarelli
          \inst{2,1}
          \orcidlink{0000-0001-9158-8580}         
          \and
          M. Bellazzini
          \inst{1}
          \orcidlink{0000-0001-8200-810X}
          \and
          A. Helmi
          \inst{3}           
          }

   \institute{INAF - Astrophysics and Space Science Observatory of 
              Bologna, Via Gobetti 93/3, 40129 Bologna, Italy\\ \email{edoardo.ceccarelli3@unibo.it}
         \and
             Department of Physics and Astronomy, University of 
             Bologna, Via Gobetti 93/2, 40129 Bologna, Italy 
         \and
             Kapteyn Astronomical Institute, University of Groningen, Landleven 12, 9747 AD Groningen, The Netherlands
             }

  \abstract
  {We present the results of the first systematic and dedicated high-resolution chemical analysis of the Thamnos substructure, a candidate relic of the process of hierarchical merging of the Milky Way. The analysis was performed in comparison with the \textit{Gaia}-Sausage-Enceladus (GSE) remnant, within the fully self-consistent and homogeneous framework established by the $\lq$A Walk on the Retrograde Side’ (WRS) project. We analysed high-resolution and high signal-to-noise ratio spectra obtained with UVES at VLT for $212$ red giant branch stars classified as candidate members of Thamnos and GSE, based on selections in the space of the integrals of motion. We derived precise abundances for $16$ atomic species. Compared to GSE, stars attributed to the Thamnos substructure are, on average, more metal-poor, yet most of them show higher [X/Fe] abundance ratios in several elements, such as Na, Mg, Al, Ca, Cu, and Zn, as well as lower [Eu/Fe]. The majority of candidate Thamnos stars show chemical signatures more consistent with the in situ Milky Way halo rather than a typical low-mass accreted dwarf galaxy. Our findings are further supported by comparisons with tailored galactic chemical evolution models, which fall short in reproducing the observed enhancement in the $\alpha$-elements, but are able to fit the more metal-poor component present in the Thamnos substructure. These results confirm a high level of contamination in the Thamnos substructure from the in situ population and to a lesser degree from GSE, while still leaving room for a genuine accreted population from a small disrupted dwarf galaxy.}

   \keywords{stars: abundances -–
             Galaxy: abundances –-
             Galaxy: formation –-
             Galaxy: halo
               }

   \maketitle


\section{Introduction}

The Milky Way (MW) likely underwent a chaotic early evolutionary path characterised by numerous mergers with smaller dwarfs, as was predicted by the standard cosmological paradigm \citep{white&frenk1991}. Disrupted galaxies leave behind debris that can be detected as phase-space overdensities in the Galactic halo with the help of dynamical information, such as the integrals of motion \citep[IoMs,][]{helmidezeeuw00}. In this context, the ESA/\textit{Gaia} mission \citep{GC16,GaiaDR3} helped the community to take a significant stride in understanding how the Galaxy formed and why it appears as it does today \citep{helmi2020,deason&belokurov2024}, and revealed the presence of numerous substructures in the dynamical spaces \citep[e.g.][]{belokurov2018,helmi2018,koppelman19,massari19,myeong19,lovdal22,oria2022,tenachi22,dodd23}. To complement this enormous dataset, chemical tagging plays a vital role, since the chemical composition of individual stars reveals features that mirror the star formation and chemical enriching history of their birthplace \citep{freeman02}. In this context, the $\lq$A Walk on the Retrograde Side' (WRS) project \citep[][Paper I hereafter]{ceccarelli2024} started an effort to build the largest high-resolution (R $> 40000$) spectroscopic dataset of stars in the retrograde halo of the MW with the goal of providing an homogeneous catalogue of chemical abundances for all the dynamically detected substructures.

In the first paper of the WRS project, we put the main focus on the chemical composition of a specific subset of retrograde substructures identified dynamically in the literature, such as ED-2 and L-RL64/Antaeus/ED-3 \citep{lovdal22,oria2022,ruiz-lara22,dodd23}, being able for the first time to fully characterise them from a chemical point of view. That study revealed that Antaeus and ED-3 have identical chemical patterns and similar IoM, suggesting a common origin. In turn, the abundance patterns of this unified system differ from that of the dominant component in the retrograde halo, the remnant of the \textit{Gaia}-Sausage-Enceladus dwarf \citep[GSE,][]{helmi2018,belokurov2018}, confirming that it is indeed an independent merging event. Also, we find negligible spread in \feh in ED-2, which may hint at this clump being the remnant of a disrupted, metal-poor stellar cluster \citep[see also][]{balbinot2023,dodd25}.

In this second paper of the WRS series, we focus on Thamnos \citep{koppelman19}, a less populated substructure compared to GSE, with generally more bounded orbits. Given its limited range in orbital energy and unusual position in the IoM spaces, Thamnos is believed to possibly be the leftover of a very early merger with a low-mass systems \citep[$M_{\star} < 5\times10^{6} \; \mathrm{M_{\odot}}$,][]{koppelman19}, although it remains uncertain whether the stars linked to this structure originate from a single merger or two separate ones, with the Thamnos 1 sub-population being on average significantly more metal-poor \citep{koppelman19,ruiz-lara22,bellazz23}. The median metallicity of Thamnos ranges from -1.4 to -1.2 dex depending on the dynamical selection of its stars, which is slightly more metal-rich than other retrograde substructures, such as Sequoia and Antaeus, hinting at some possible contamination by stars with different origin \citep{lovdal22,ruiz-lara22,bellazz23,dodd23,mori2025}. Indeed, the very limited chemical information available for the Thamnos substructure seems to suggest a significant amount of contamination from the in situ MW and possibly also GSE, as some of its stars have high \afeh ratios \citep{monty2020,dodd23,horta23,zhang2024}. This hypothesis has been further strengthened by the application of substructure identification to cosmological hydrodynamical simulations, which show that the IoM region occupied by Thamnos can be largely populated by in situ stars and/or by stars from different progenitors \citep{thomas25}. However, \citet{dodd25SFH} find via a colour-magnitude diagram (CMD)-fitting technique \citep{gallart24} that the age-metallicity distribution at low metallicity $(\feh < -1.5 \dex)$ for stars dynamically linked to the Thamnos substructure is statistically different to that of these possible contaminants. These authors argue that it is consistent with the presence of an accretion event that took place earlier than GSE, and they also conclude that there is significant contamination (at least $50\%$) especially at higher metallicity. In this work we aim to provide the very first detailed high-resolution chemical characterisation of a large sample of dynamically selected stars from the Thamnos substructure in the self-consistent framework of the WRS project, with the goal of shedding light on the nature of this object. 
 
The paper is organised as follows. In Section~\ref{sec:dataset} we present the observational strategy and the dataset used in this work. In Section~\ref{sec:orbital_parameters} and Section~\ref{sec:chemical_abu} we describe the methods employed to carry the chemo-dynamical analysis and present the results. In Section~\ref{sec:GCE_models} we compare the abundances with predictions from galactic chemical evolution (GCE) models. Finally, in Section~\ref{sec:discussion} we discuss the results and provide a summary of the key findings of our study. In what follows, we refer to Thamnos and the Thamnos substructure indistinctly. As we progress in our analysis and especially in later sections,  we refer to the $\lq$true Thamnos' as a putative accreted component distinct from GSE and the in situ population.

\section{Dataset and observations}
\label{sec:dataset}

\begin{figure}[!th]
\centering
\includegraphics[width=0.4\textwidth]{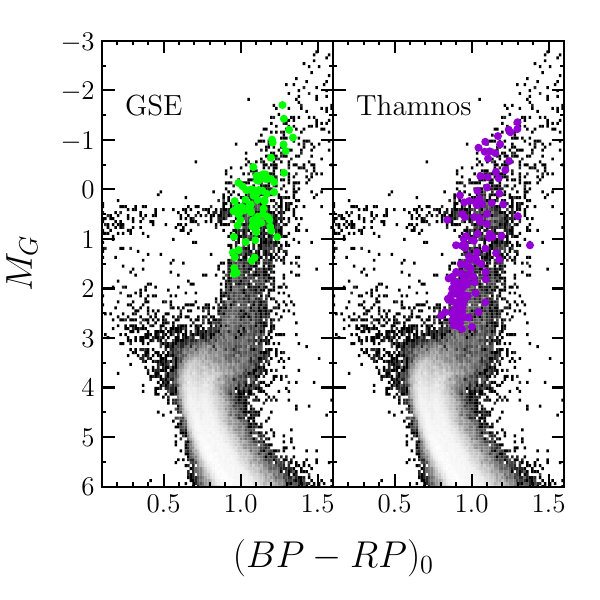} 
\caption{CMD of stars dynamically associated with GSE and Thamnos (filled green and purple symbols, respectively). In the background a density map of nearby MW halo stars ($d < 2$ kpc) is plotted.}
\label{fig:cmd}%
\end{figure}

We used the catalogue of accreted stars based on the third data release of \textit{Gaia} \citep[DR3,][]{GaiaDR3} provided in \citet{dodd23}, selectively focusing on bright ($G < 15$) red giant branch (RGB) stars linked to Thamnos and GSE (see their Section 2 for all the quality cuts applied to \textit{Gaia} data). The final sample comprises $142$ stars associated with Thamnos and $78$ to GSE. In Fig. \ref{fig:cmd} we show the CMD of selected targets superimposed to a sample of the full MW halo within $2$ kpc. Magnitudes have been corrected taking $E(B-V)$ values from the \texttt{EXPLORE}\footnote{\url{https://explore-platform.eu/}} webpage.

\begin{figure}[!th]
\centering
\includegraphics[width=0.4\textwidth]{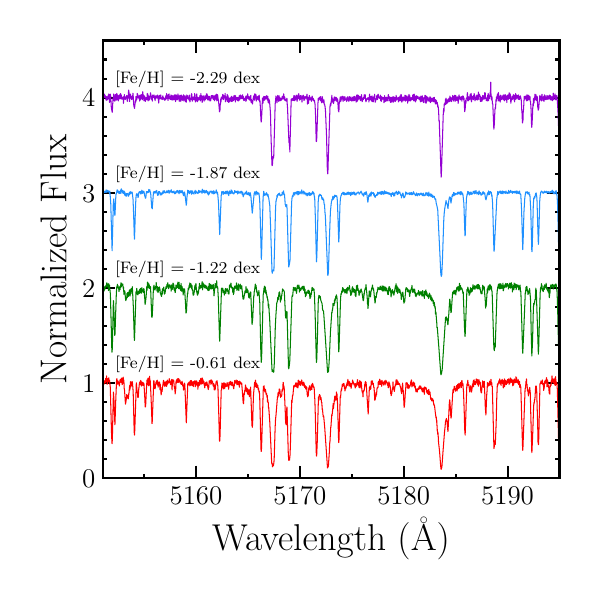} 
\caption{UVES spectra of four target stars with similar atmospheric parameters and different metallicity. The spectra have been arbitrarily shifted vertically for plotting purposes.}
\label{fig:spectra}%
\end{figure} 
\begin{table*}
\caption{Information for some representative targets.}        
\label{tab:obs}      
\centering          
\begin{tabular}{ccccccccc}  
\hline 
\hline      
ID \textit{Gaia} DR3 & RA & Dec & $E(B-V)$ & $G$ & $t_{\mathrm{exp}}$ & Airmass & S/N & S/N \\ 
 & (deg) & (deg) & (mag) & (mag) & (s) &  & ($3900$ \r{A}) & ($5800$ \r{A})    \\ 
\hline 
5138126933062532352      & 30.8293700 & -18.833342 & 0.02 & 10.5038    & 200    &      1.727             &      27       &      71       \\
3541053961204824832      & 172.587726 & -22.548268 & 0.04 & 11.2243    & 400    &      1.298             &      20       &      48       \\
5828822717270400128      & 243.864451 & -64.377047 & 0.15 & 12.4347    & 1000   &      1.302             &      22       &      62       \\
6435410912786146816      & 288.023829 & -65.008052 & 0.06 & 13.4251    & 1600   &      1.195             &      29       &      75       \\
6794113593364452608      & 312.424452 & -31.081279 & 0.06 & 13.8453    & 3000   &      1.260             &      32       &      78       \\
\hline                  
\end{tabular}
\tablefoot{ID and co-ordinates from \textit{Gaia} DR3, colour excess, apparent $G$ magnitude, exposure time ($t_{\mathrm{exp}}$), airmass, and S/N at the centre of the blue and red arms of UVES for some of the observed target spectra. 
}
\end{table*}
 
The high-resolution spectra of the targets were collected with the optical spectrograph UVES \citep{dekker2000} at the Very Large Telescope of ESO under the programs 0112.B-0236 (P.I.: E. Ceccarelli) and 0113.B-0196 (P.I.: E. Ceccarelli). Observations were taken with UVES in Dichroic mode using the Dic 1 Blue Arm CD2 390 (3300 - 4500 \r{A}) and the Dic 1 Red Arm CD3 580 (4800 - 6800 \r{A}) and adopting the 1x12 slit (R $= 40000$). The average signal-to-noise ratio (S/N) is typically higher than $25$ for the blue arm and $50$ for the red arm. The obtained target spectra were reduced with the dedicated ESO pipeline\footnote{\url{https://www.eso.org/sci/software/pipelines/}}. We provide in Fig. \ref{fig:spectra} the spectra of several targets with similar atmospheric parameters (i.e. effective temperature and surface gravity) but different [Fe/H]. Main information on representative target stars and the observations can be found in Table \ref{tab:obs}.
      
\section{Orbital parameters}
\label{sec:orbital_parameters}

\begin{figure*}[!th]
\centering
\begin{minipage}{0.33\textwidth}
\centering
\includegraphics[width=1.0\textwidth]{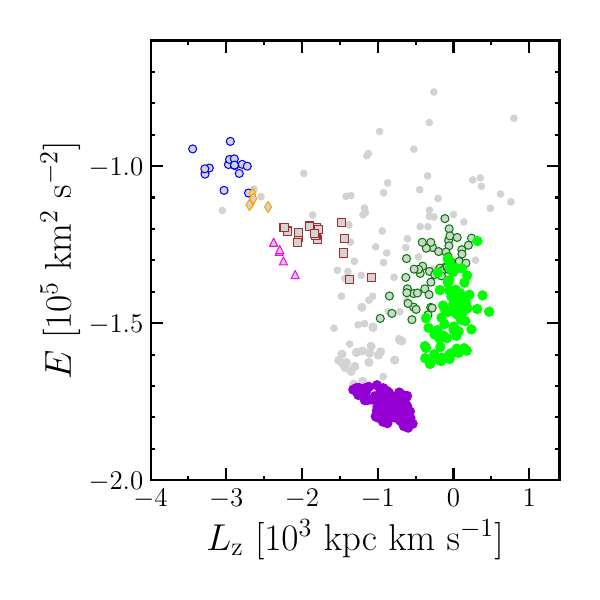} 
\end{minipage}
\begin{minipage}{0.33\textwidth}
\centering
\includegraphics[width=1.0\textwidth]{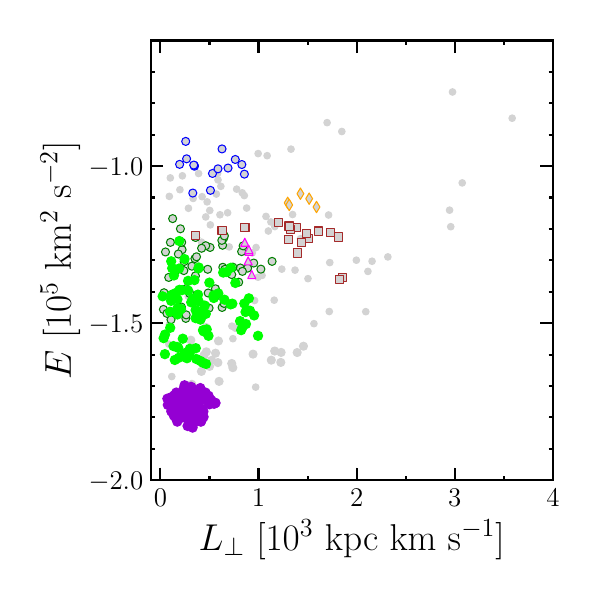}
\end{minipage}
\begin{minipage}{0.33\textwidth}
\centering
\includegraphics[width=1.0\textwidth]{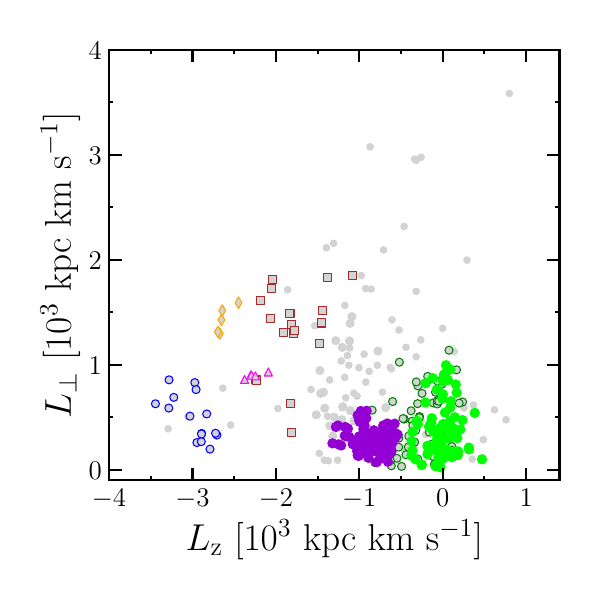}
\end{minipage}   
\caption{Distribution of the observed stars in $E$ - $L_{\mathrm{z}}$ - $L_{\perp}$ spaces. Stars associated with Thamnos and GSE by \citet{dodd23} are plotted in purple and green, respectively. Stars from \citetalias{ceccarelli2024} are reported in background as grey points. The colour coding of the border of these points reflect the association provided in \citetalias{ceccarelli2024}: dark green for GSE, brown for Sequoia, blue for Antaeus, magenta for ED-2, orange for ED-3, and grey for stars that are not associated with any substructure.}
\label{fig:op}%
\end{figure*} 
%
\begin{table*}
\caption{Dynamical information for selected targets (extract).}          
\label{tab:op}      
\centering          
\begin{tabular}{cccccccccc}  
\hline 
\hline 
ID \textit{Gaia} DR3 & E & $\sigma$(E) & $L_{\mathrm{z}}$ & $\sigma$($L_{\mathrm{z}}$) & $L_{\mathrm{perp}}$ & $\sigma$($L_{\mathrm{perp}}$) & $V_{\mathrm{los}}$ & $\sigma$ ($V_{\mathrm{los}}$) & Binary Star\\ 
 & \multicolumn{2}{c}{(x$10^{5}$ $\mathrm{km^{2}}$ $\mathrm{s^{-2}}$)} & \multicolumn{2}{c}{(kpc km $\mathrm{s^{-1}}$)} & \multicolumn{2}{c}{(kpc km $\mathrm{s^{-1}}$)} & \multicolumn{2}{c}{(km $\mathrm{s^{-1}}$)} & \\ 
\hline 
5138126933062532352      &      -1.599   &      0.010    &      -203     &         53     &        46     &        22     &      -52.5    &      0.5          &  no \\
3541053961204824832      &      -1.323   &      0.047    &      -46      &         83     &       391     &        36     &      133.3    &      0.4          &  no         \\
5828822717270400128      &      -1.790   &      0.009    &      -728     &         37     &       185     &        10     &      315.7    &      0.4          &  no         \\
...      & ...   & ...   & ...   & ...   & ...   & ...   & ...   & ...   & ...      \\
\hline                  
\end{tabular}
\tablefoot{ID from \textit{Gaia} DR3, orbital energy, angular momentum along the $z$ axis, perpendicular angular momentum, and line-of-sight velocity with uncertainties. The entire table is available at the CDS. 
}
\end{table*}

A detailed description of the methods used to study the orbits of target stars is presented in \citetalias{ceccarelli2024}, and we refer to Section 3 of the quoted paper for the interested reader. A brief description can be found in Appendix \ref{app:op}. The median orbital parameters for each target, together with the line-of-sight velocity, are listed in Table \ref{tab:op}.

In Fig. \ref{fig:op} we show the position of our targets superimposed with stars from \citetalias{ceccarelli2024}. We remind the reader that in \citetalias{ceccarelli2024}, we selected retrograde stars in the MW halo by imposing a cut in the linear velocity in the sky plane ($V_{\mathrm{T}}>400$ km s$^{-1}$, see Sections 2.1 and 7.1 in \citetalias{ceccarelli2024} for a complete discussion). This criterion prevented us to select either stars with low orbital energy ($E \le -1.7 \times 10^{5}$  $\mathrm{km^{2}}$ $\mathrm{s^{-2}}$) moving on retrograde orbits ($L_{\mathrm{z}} \sim -1000$ kpc km $\mathrm{s^{-1}}$) and stars with slightly higher orbital energy ($E \le -1.5 \times 10^{5}$  $\mathrm{km^{2}}$ $\mathrm{s^{-2}}$) and almost null net rotation, that is where Thamnos and the bulk population of GSE are located, respectively. As is evident from Fig. \ref{fig:op}, we are able in this work to fully sample these two substructures.

\section{Chemical abundances}
\label{sec:chemical_abu}
\begin{figure}[!th]
\centering
\includegraphics[width=0.45\textwidth]{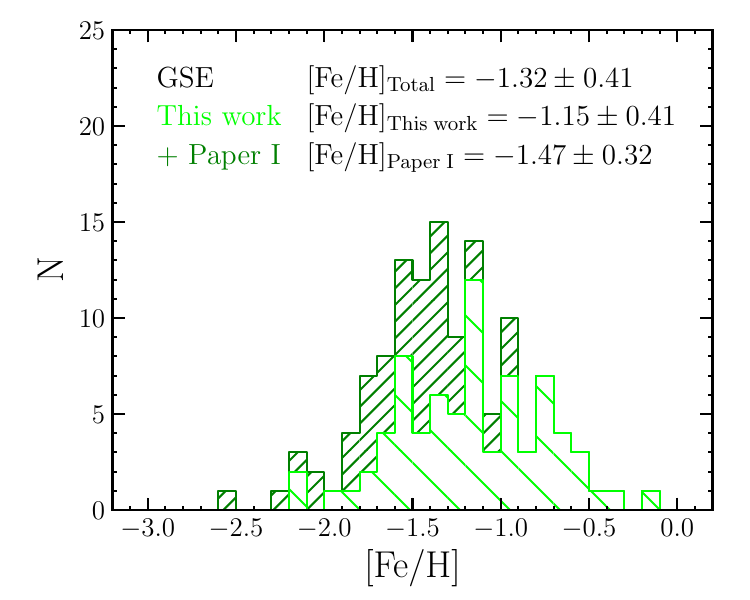} 
\includegraphics[width=0.45\textwidth]{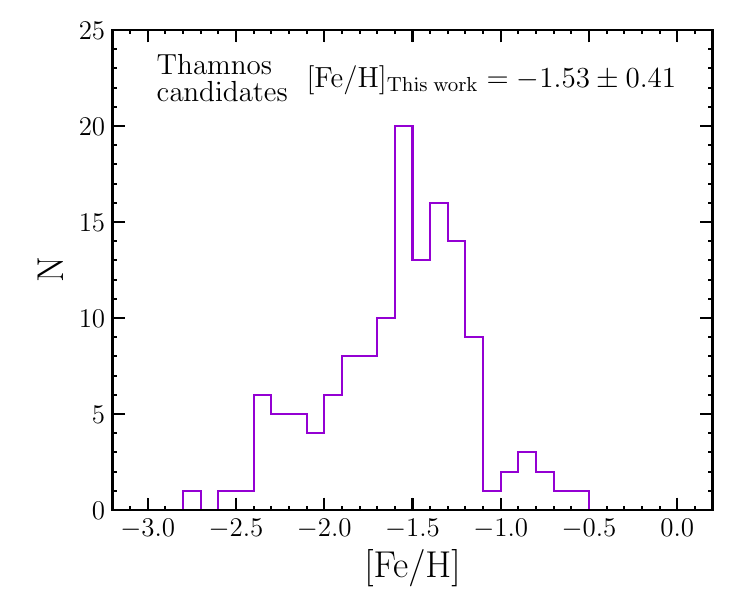}   
\caption{Metallicity distributions of the GSE (top panel) and Thamnos (bottom panel) coming from neutral Fe lines. The median metallicity of the distributions and the standard deviation are also reported in each panel. For GSE we report the MDF derived in this work (green) alongside the stacked MDF using also stars from \citetalias{ceccarelli2024} (dark green). We note that the shift in metallicity we find between the two samples in GSE might be due to a gradient in the progenitor reflected by the position of its stars in the $E$ - $L_{\mathrm{z}}$ plane (see discussion in the text).}
\label{fig:mdf}%
\end{figure} 
\begin{figure}
\centering
\includegraphics[width=0.45\textwidth]{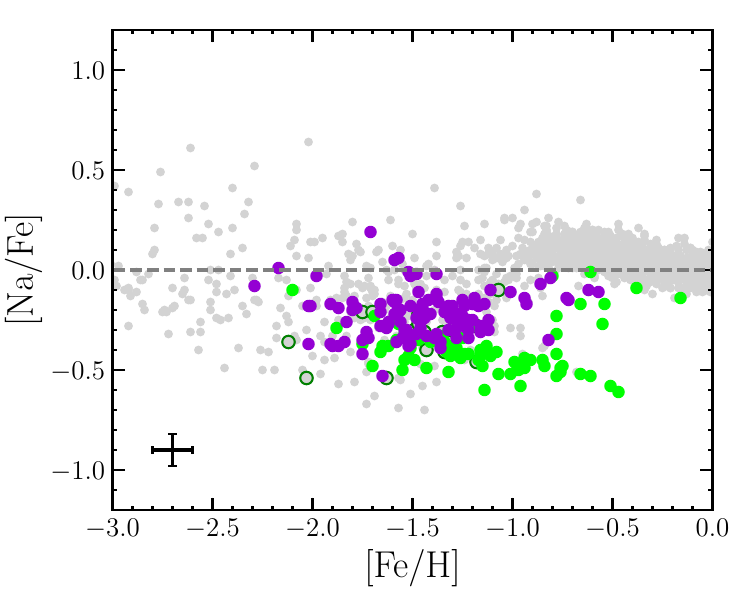} 
\includegraphics[width=0.45\textwidth]{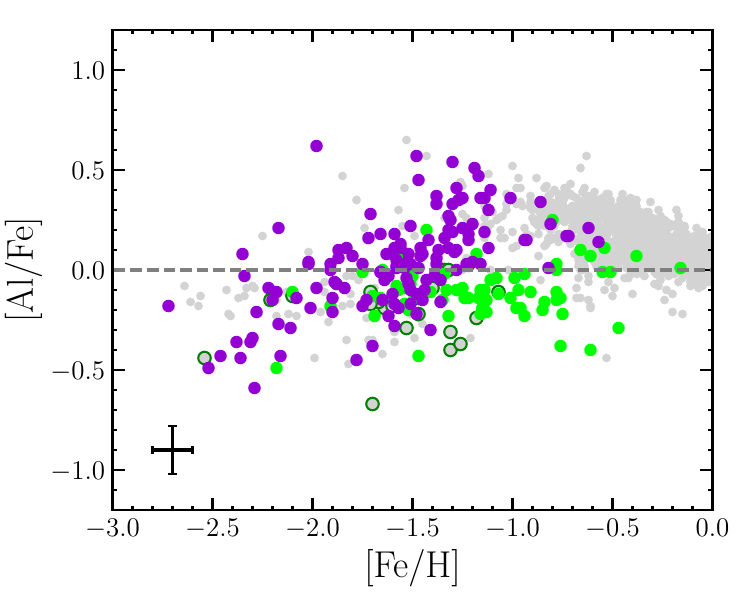}  
\caption{Abundance ratios of the light elements Na and Al for the selected stars corrected for NLTE effects. The colour coding is the same as in Fig. \ref{fig:op}. In the lower left corner of each panel, we report typical uncertainties for the abundance ratios. Literature abundances for MW stars are taken from \citet{edvardsson93,fulbright2000,stephens2002,gratton03,reddy2003,reddy06,barklem05,bensby05,bensby14,roederer2014}, \citet{reggiani2017}, and \citetalias{ceccarelli2024}. Grey points with dark green borders are GSE stars from \citetalias{ceccarelli2024}.}
\label{fig:abu_light}%
\end{figure}
\begin{figure}
\centering
\includegraphics[width=0.45\textwidth]{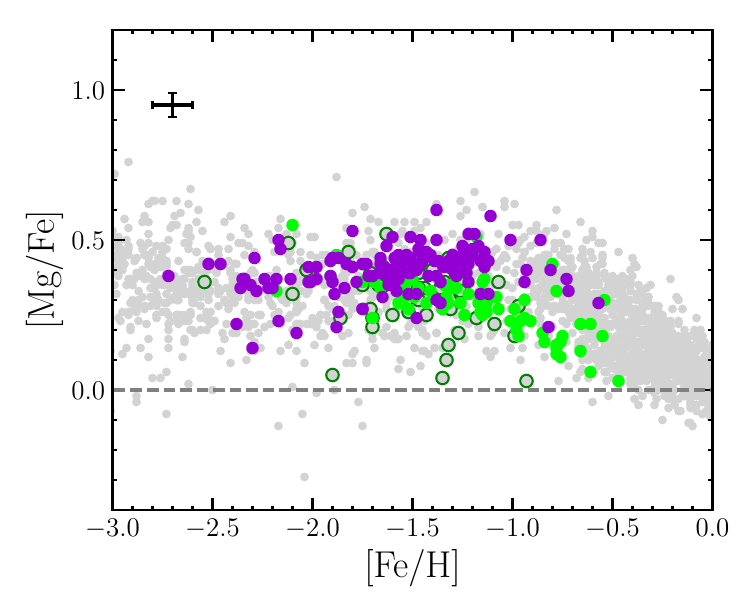} 
\includegraphics[width=0.45\textwidth]{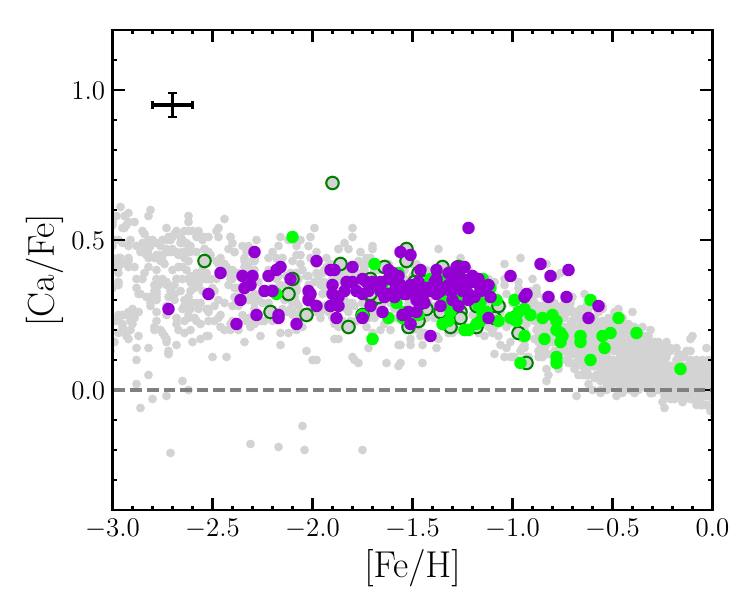}  
\includegraphics[width=0.45\textwidth]{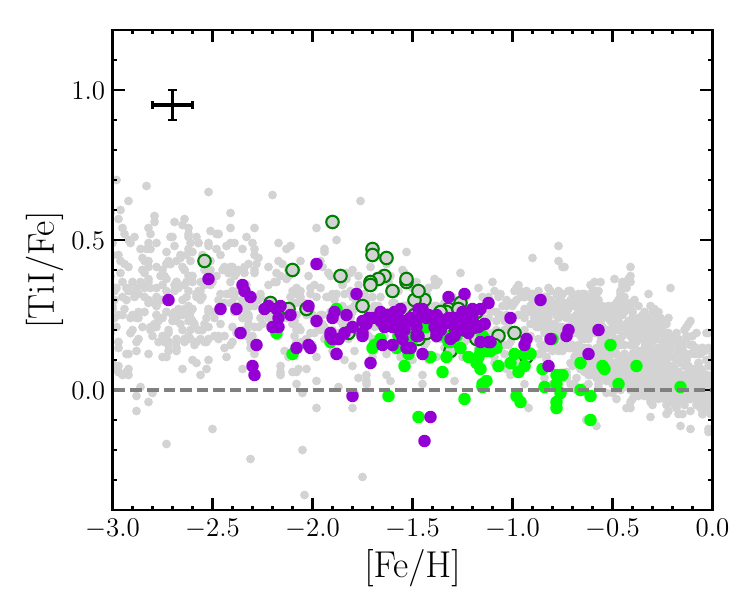}       \caption{Same as Fig. \ref{fig:abu_light}, but for Mg, Ca, and Ti.} 
\label{fig:abu_alpha}%
\end{figure}
\begin{figure*}
\centering
\begin{minipage}{0.45\textwidth}
\centering        
\includegraphics[width=1.0\textwidth]{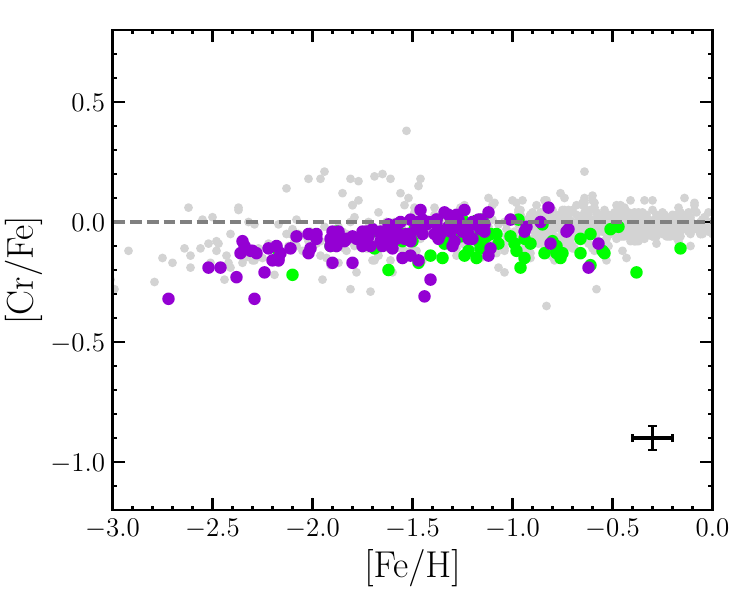} 
\includegraphics[width=1.0\textwidth]{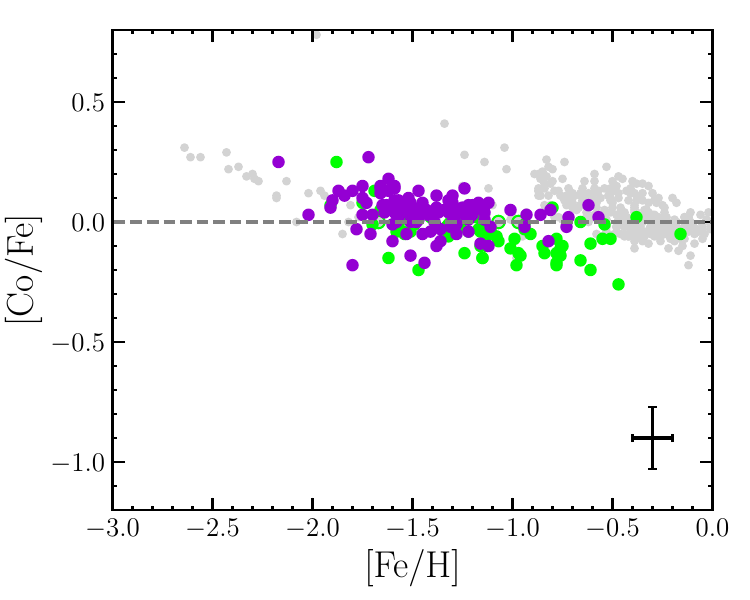}
\includegraphics[width=1.0\textwidth]{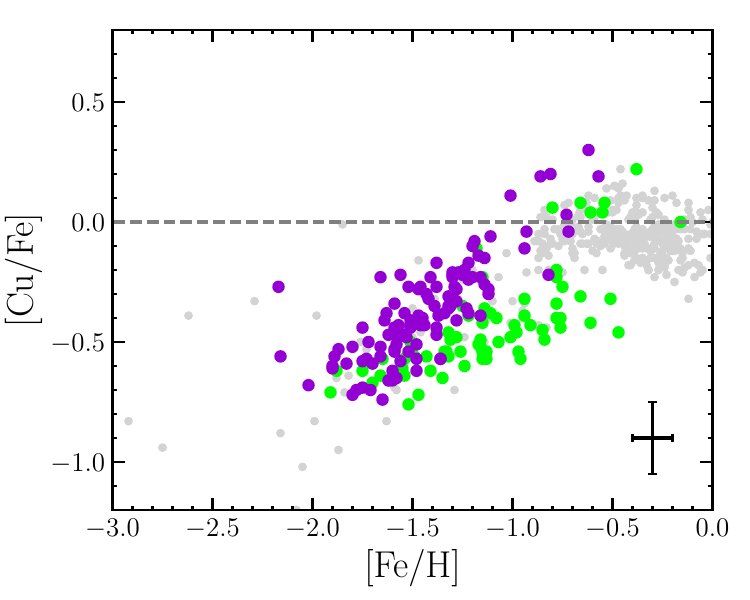}        
\end{minipage}
\begin{minipage}{0.45\textwidth}
\centering
\includegraphics[width=1.0\textwidth]{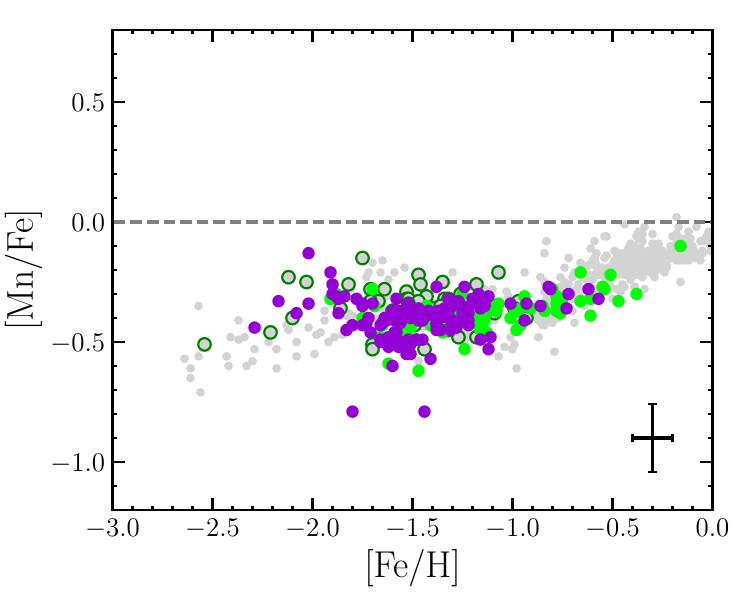}
\includegraphics[width=1.0\textwidth]{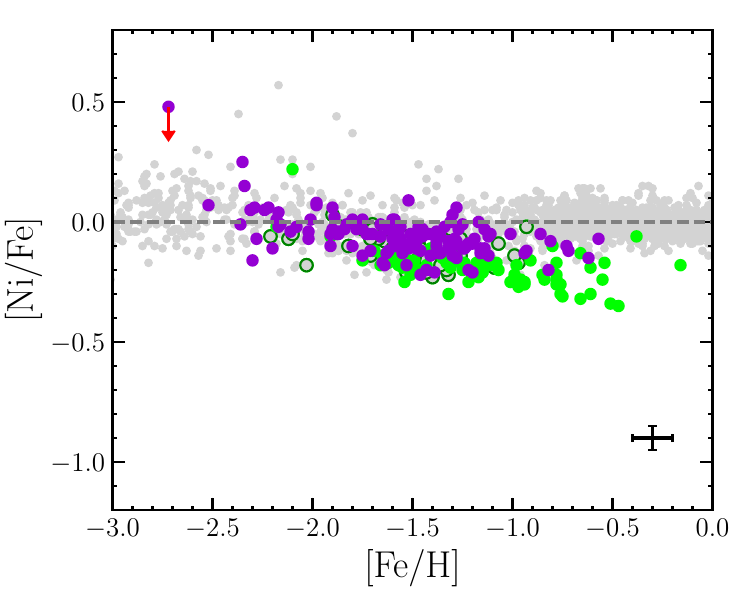}
\includegraphics[width=1.0\textwidth]{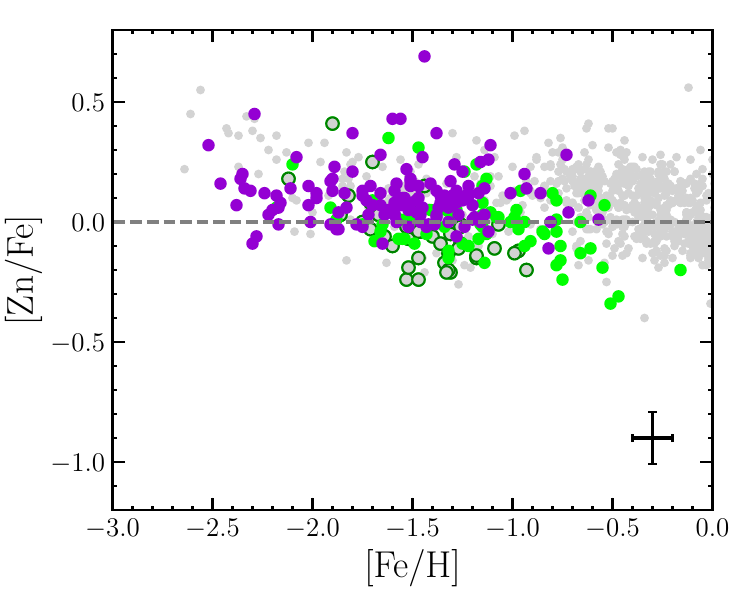}        
\end{minipage}  
\caption{Same as Fig. \ref{fig:abu_light}, but for Cr, Mn, Co, Ni, Cu, and Zn. Upper limits are plotted as red arrows.} 
\label{fig:abu_ironpeak}%
\end{figure*}
\begin{figure*}
\centering
\begin{minipage}{0.45\textwidth}
\centering
\includegraphics[width=1.0\textwidth]{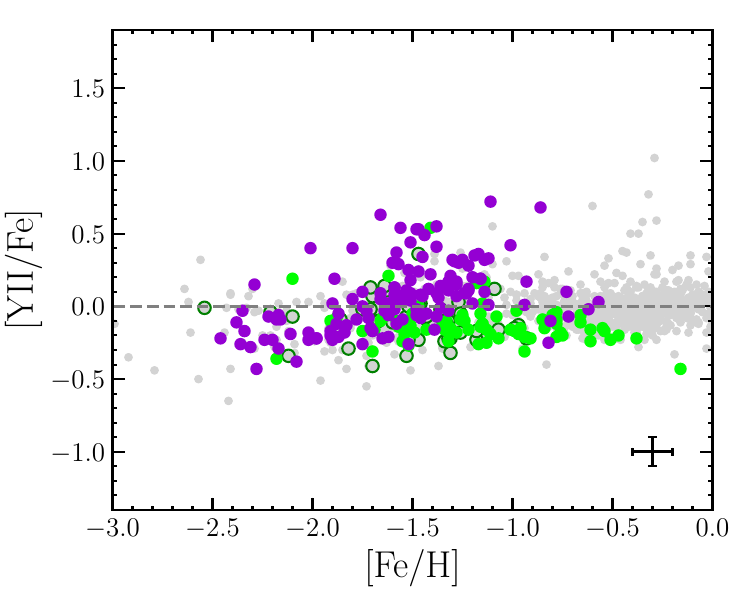} 
\includegraphics[width=1.0\textwidth]{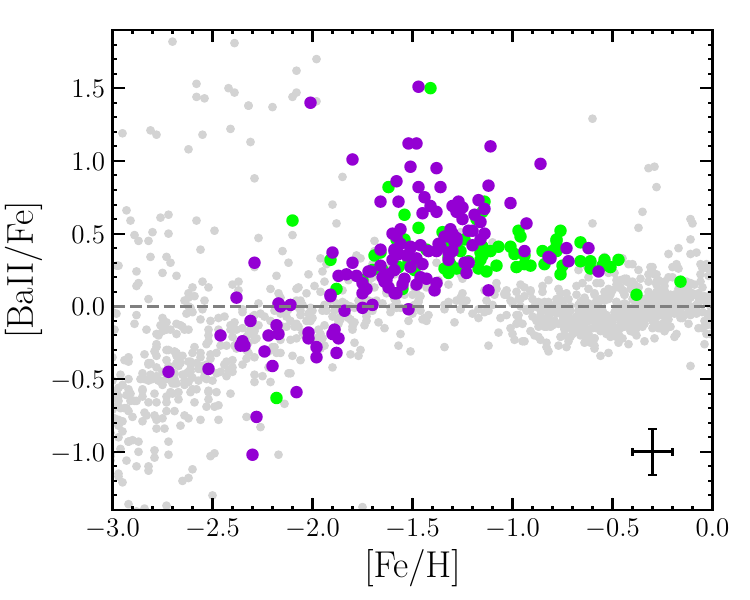}
\end{minipage}
\begin{minipage}{0.45\textwidth}
\centering
\includegraphics[width=1.0\textwidth]{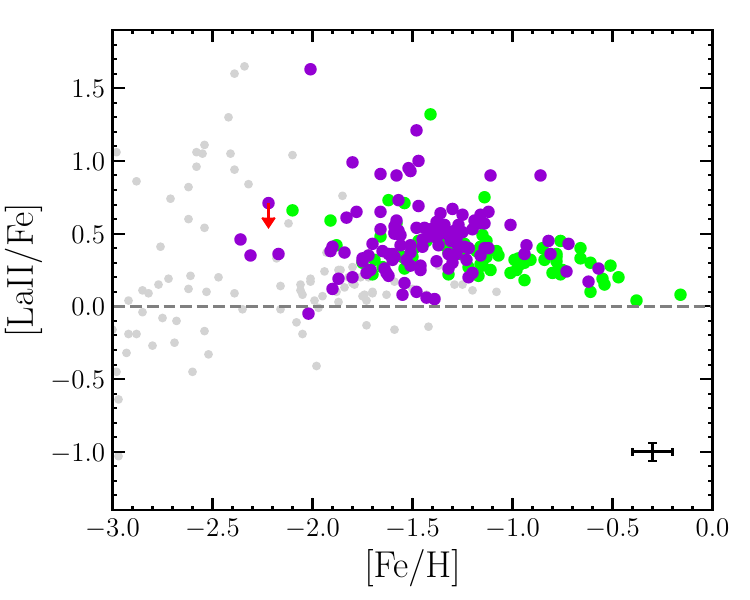}
\includegraphics[width=1.0\textwidth]{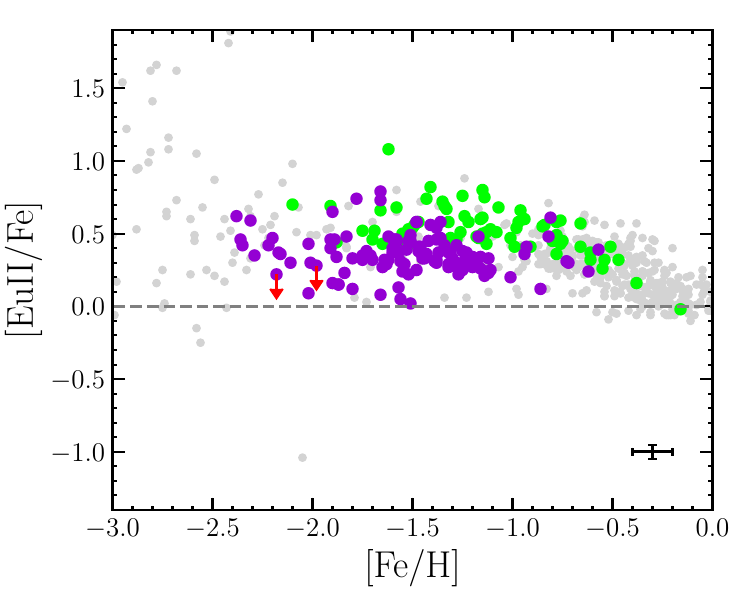}       
\end{minipage}  
\caption{Same as Fig. \ref{fig:abu_light} but for Y, La, Ba, and Eu. Upper limits are plotted as red arrows.}
\label{fig:abu_neutroncapture}%
\end{figure*}
\begin{figure}
\centering
\includegraphics[width=0.45\textwidth]{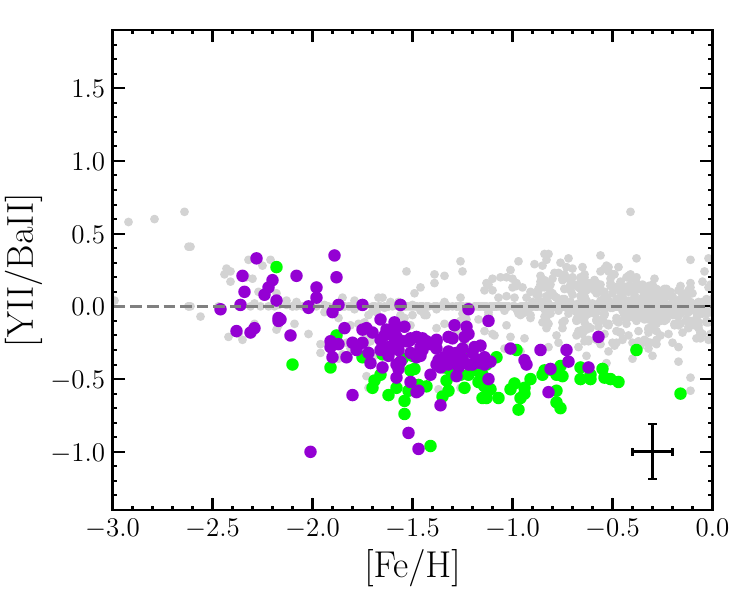}  
\includegraphics[width=0.45\textwidth]{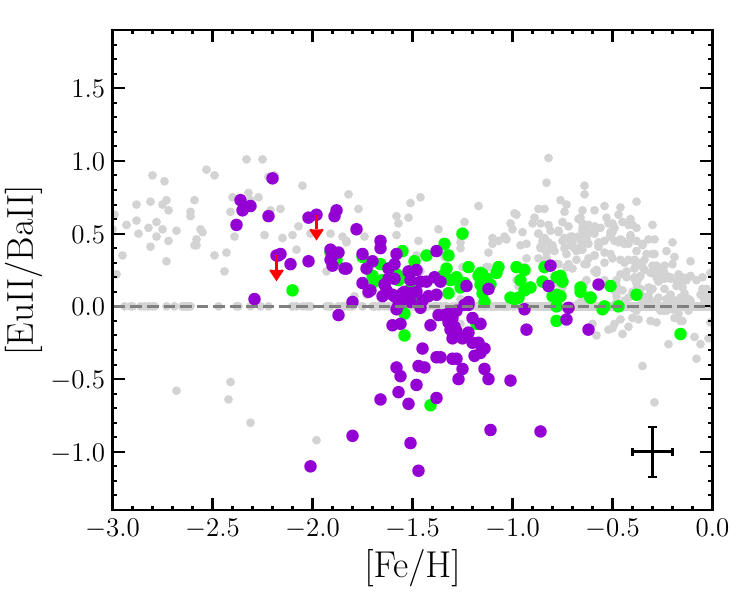}      
\includegraphics[width=0.45\textwidth]{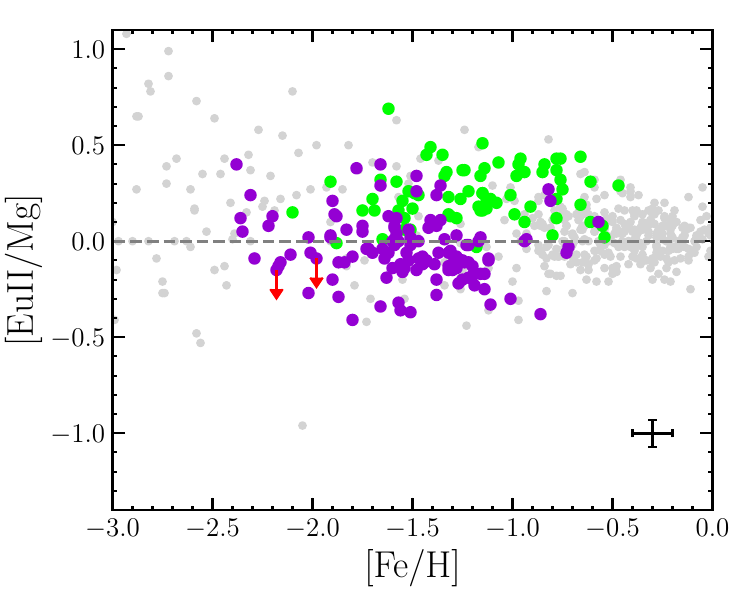}  
\caption{Same as Fig. \ref{fig:abu_light} but for the [YII/BaII], [EuII/BaII], and [EuII/Mg] ratios. Upper limits are plotted as red arrows.}
\label{fig:abu_neutroncapture2}%
\end{figure}
\begin{table}
\caption{Stellar parameters for the selected targets (extract).}          
\label{tab:parameters}      
\centering          
\begin{tabular}{ccccc}  
\hline
\hline
ID \textit{Gaia} DR3 & $T_{\mathrm{eff}}$ & log $g$ & $v_{\mathrm{t}}$   \\ 
 & (K) & (dex) & (km $\mathrm{s^{-1}}$)  \\ 
\hline 
5138126933062532352      &      5157     &      2.44     &      1.5      \\
3541053961204824832      &      4864     &      2.03     &      1.4      \\
5828822717270400128      &      5183     &      2.35     &      1.9      \\
...      & ...   & ...   & ...   \\
\hline                  
\end{tabular}
\tablefoot{ID from \textit{Gaia} DR3, effective temperature ($T_{\mathrm{eff}}$), surface gravity (log $g$), and microturbulent velocity ($v_{\mathrm{t}}$). Typical uncertainties on $T_{\mathrm{eff}}$, log $g$, and $v_{\mathrm{t}}$ are of the order of 100 K, 0.1 dex and 0.2 km $\mathrm{s^{-1}}$. The entire table is available at the CDS. 
}
\end{table}
\begin{table*}
\caption{Chemical abundances for the light, $\alpha$, iron-peak and neutron capture elements (extract).}\label{tab:abundances}
\centering
\begin{tabular}{cccccc} 
\hline\hline             
ID \textit{Gaia} DR3 & [Fe/H] & [Na/Fe] & [Al/Fe] & [Mg/Fe] & [Ca/Fe] \\ 
\hline 
5138126933062532352      &      -0.78 $\pm$ 0.10 &       -0.23 $\pm$ 0.05        & 0.00 $\pm$ 0.12  &      0.14 $\pm$ 0.02 & 0.09 $\pm$ 0.03    \\
3541053961204824832      &      -1.34 $\pm$ 0.11 &       -0.36 $\pm$ 0.08        & -0.02 $\pm$ 0.13  &      0.35 $\pm$ 0.03 & 0.31 $\pm$ 0.04    \\
5828822717270400128      &      -2.31 $\pm$ 0.10 &       -        & -0.36 $\pm$ 0.12  &      0.35 $\pm$ 0.04 & 0.35 $\pm$ 0.04  \\
...      & ...   & ...   & ...   & ...   & ...        \\
   \hline             
[TiI/Fe] & [ScII/Fe]  & [Cr/Fe]  & [Mn/Fe] & [Ni/Fe] & [Cu/Fe]  \\ 
\hline 
-0.06 $\pm$ 0.05   &      0.10 $\pm$ 0.07     & -0.13  $\pm$ 0.03              & -0.31 $\pm$ 0.14      & -0.17 $\pm$ 0.04       & -0.20 $\pm$ 0.16                    \\
0.18 $\pm$ 0.06    &      0.11 $\pm$ 0.08     & -0.09  $\pm$ 0.03              & -0.39 $\pm$ 0.15      & -0.15 $\pm$ 0.04       & -0.54 $\pm$ 0.16                    \\
0.31 $\pm$ 0.04    &       0.11 $\pm$ 0.07     & -0.12  $\pm$ 0.03              & -      & 0.05 $\pm$ 0.05         & -                        \\                 
...           & ...   & ...   & ...   & ...   & ...                  \\
   \hline             
[Zn/Fe] &  [YII/Fe]  & [BaII/Fe]  & [LaII/Fe] & [EuII/Fe] & Progenitor \\ 
\hline 
-0.04 $\pm$ 0.12                            & -0.12 $\pm$ 0.10      & 0.46 $\pm$ 0.17       & 0.36 $\pm$ 0.06             & 0.36 $\pm$ 0.05        & GSE \\
-0.02 $\pm$ 0.13                            & -0.16 $\pm$ 0.10      & 0.26 $\pm$ 0.17       & 0.42 $\pm$ 0.06             & 0.69 $\pm$ 0.05        & GSE \\
0.13 $\pm$ 0.06                            & -0.28 $\pm$ 0.10      & -0.10 $\pm$ 0.17      & 0.35 $\pm$ 0.06             & 0.59 $\pm$ 0.05        & Thamnos \\
...              & ...           & ...   & ...   & ... & ...                   \\
\hline
\end{tabular}
\tablefoot{In the last column we report the dynamical association with the former progenitor by \citet{dodd23}. The entire table is available at the CDS. 
}  
\end{table*}

To erase the chance that some offset is introduced in the final results due to differences in the assumptions in the chemical analysis, we stuck to the procedure followed in \citetalias{ceccarelli2024}. A brief summary of the method is reported in Appendix \ref{app:ap} - \ref{app:ca}. 

Among the stars analysed in this work, we find four showing significant C enhancement, resulting in spectra that are heavily contaminated by molecular absorption features. Additionally, four stars show high Li abundances. Of these eight chemically peculiar stars, five are associated with Thamnos, and three with GSE. Since the primary objective of this work is to interpret abundance patterns as tracers of the distinct chemical enrichment histories of the progenitors, we neglected stars that behave in a peculiar way compared to the bulk of the populations under study. Carbon and lithium enrichment are commonly linked to binary evolution, implying that the observed abundances may no longer reflect the original chemical signature of the formation environment but rather the influence of the co-evolution with the companion star. For this reason, we excluded these stars from the subsequent analysis. Their properties will be explored in detail in a forthcoming, dedicated study.

In the end, we derive abundances for 137 and 75 stars associated with Thamnos and GSE, respectively. The final atmospheric parameters are listed in Table \ref{tab:parameters}. Values of the abundance ratios of light, $\alpha$, iron-peak and neutron capture elements are reported in Table \ref{tab:abundances}. In the following we discuss results obtained from the analysis.

\subsection{Metallicity distributions}
\label{subsec:MDFs}

We used the Fe abundance derived from approximately $100$ neutral lines per star as a proxy for metallicity. In Fig. \ref{fig:mdf} we show the metallicity distribution functions (MDFs) of GSE and Thamnos (top and bottom panels, respectively). The median and standard deviation are reported in each panel. We also display for GSE the stacked histogram obtained adding the sample from \citetalias{ceccarelli2024} (dark green line). We note that the two GSE samples analysed in \citetalias{ceccarelli2024} and in the present work are entirely independent, with no stars in common. We find that the MDF of the GSE stars analysed in this work is, on average, $0.3 \dex$ more metal-rich than that reported in \citetalias{ceccarelli2024}. This is somehow expected, as the selection criteria adopted in \citetalias{ceccarelli2024} preferentially target GSE stars moving on higher-energy and more retrograde orbits (see left panel of Fig. \ref{fig:op}), which are likely to represent the earliest stars stripped during the merging event and to originate from the outer regions of the progenitor \citep{koppelman2020,amarante2022,skuladottir25}. Thus, if a negative metallicity gradient was in place in GSE, as is observed for surviving dwarf galaxies \citep[see e.g.][]{tolstoy09}, stars from \citetalias{ceccarelli2024} should be on average more metal-poor than the bulk population of GSE. Taken together, these considerations suggest that the most representative picture of the GSE progenitor is obtained by combining the two samples, with the present work tracing the bulk of the system (due to the dynamical selection of \citet{dodd23}, which prefers purity over completeness), and \citetalias{ceccarelli2024} preferentially probing stars presumably stripped from the outskirts of the progenitor galaxy. The peak and shape of the overall GSE MDF are consistent with previous literature investigations \citep[e.g.][]{helmi2018,vincenzo2019,naidu20,myeong22,dodd25SFH}. For Thamnos candidate members, we find that the MDF is slightly more metal-poor than that of GSE, with a peak at $\feh \sim -1.5 \dex$, and a slightly more pronounced tail at low metallicity, similar to that presented in \citet{dodd25SFH}. 

Given that stars in Thamnos display low orbital energy and retrograde motion, a region of the IoM space typically dominated by in situ stars \citep[i.e. the Aurora population,][]{belokurov22}, we expect some level of contamination from such a stellar population \citep[see also][]{dodd25SFH}. 
 Also, if the Aurora population is indeed the dominant source of contamination, it could also explain the sharp decline in the MDF observed at $\feh \sim -1.3 \dex$. The earlier mentioned excess of stars with $\feh < -2.0 \dex$ relative to GSE, is consistent with findings by \citet{dodd25SFH}, and may signal the presence of the residual stellar debris from the $\lq$true Thamnos' progenitor. In the following, we discuss the detailed chemical abundances of several elements that will help in disentangling the contribution of in situ contamination from the real accreted population in the Thamnos substructure.

\subsection{Abundances of light elements}

In the low-metallicity regime, Na and Al are primarily synthesised in massive stars that explode as core-collapse supernovae (CCSNe), whereas at intermediate metallicities they are also produced in non-negligible amounts in asymptotic giant branch (AGB) stars. For both elements, the production depends on the metallicity as it stems from a neutron excess generated during the CNO cycle \citep{kobayashi2009,nomoto2013}. 

In Fig. \ref{fig:abu_light} we present the abundance ratios of light elements Na and Al as a function with \feh for Thamnos and GSE (purple and green points, respectively). Specifically, Na has been measured through EW using the doublet at 5680 \r{A}, while abundances for Al have been measured via spectral synthesis of the doublets at 3944 - 3962 \r{A} and 6696 - 6698 \r{A}. Abundances of Na and Al have been corrected for NLTE effects interpolating the grids provided by \citet{lind22}.
 
As is evident from the top panel of Fig. \ref{fig:abu_light}, both Thamnos and GSE show tight patterns in [Na/Fe], yet they differ on average by 0.2 dex at fixed metallicity. In particular, Thamnos shows a flat trend at [Na/Fe] $\sim -0.3 \dex$ up to $\feh \sim -0.5 \dex$, while GSE is depleted, with [Na/Fe] $\sim -0.5 \dex$. 

A different behaviour between the two substructures can be seen in Al abundances, too. Indeed, GSE stars follow a flat trend with [Al/Fe] $\sim -0.2\dex$ across the entire metallicity range, in agreement with results from \citetalias{ceccarelli2024} (grey points with dark green borders). In contrast, Thamnos stars show a clear increase in [Al/Fe] with metallicity: their values are comparable to those of GSE up to $\feh \sim -1.7 \dex$, beyond which they rise steadily, reaching enhancements of of $\sim 0.4 \dex$ at $\feh \sim -1.0$ dex and overlapping with the trend described by the in situ population. This behaviour is typical of an environment that efficiently undergoes rapid chemical self-enrichment, allowing the [Al/Fe] to grow before the onset of SN Ia, as the yield of Al strongly depends on the metallicity of the progenitor star \citep{hawkins2015,hayes2018,kobayashi2020}. On the contrary, the star formation in GSE-like systems is not efficient enough to produce such a trend.
 
\subsection{Abundances of $\alpha$-elements}
\label{subsec:alphaFe}

It is well established that the \afeh is an effective tracer of the star formation timescales and efficiency of a given environment \citep[time-delay model, see e.g.][]{Matteucci12}. The interstellar medium is enriched on short timescales with both $\alpha$-elements and Fe due to the explosion of CCSNe. However, Fe is also significantly produced by Type Ia supernovae (SN Ia), which originate from the explosions of white dwarfs after binary interactions in low-mass stellar systems \citep{kobayashi2020}. Given the delayed onset of SN Ia, systems with inefficient star formation timescales tend to exhibit lower \afeh ratios \citep{Matteucci90}.

In this work, Mg abundances are derived from the 5528 \r{A} and 5711 \r{A} lines. To ensure consistency with the results presented in \citetalias{ceccarelli2024}, where only the Mg b triplet lines were available, we additionally derived Mg abundances using this feature for the current dataset. We find a systematic offset of $0.12 \dex$ ($\sigma = 0.04 \dex$) between the two sets of measurements, with the Mg b triplet yielding lower abundances. Therefore, to enable a homogeneous comparison, the Mg abundances from \citetalias{ceccarelli2024} presented in top panel of Fig. \ref{fig:abu_alpha} (grey points with dark green borders) have been rescaled by this offset. We note that the GSE stars analysed in this work follow a well-defined [Mg/Fe] trend that closely aligns with the main distribution reported by \citetalias{ceccarelli2024}. However, we also identify a few stars from the \citetalias{ceccarelli2024} sample exhibiting [Mg/Fe] $< 0.2 \dex$ at $\feh < -1.3 \dex$. These outliers may originate from lower-mass accreted systems and could have been inadvertently included in the GSE sample due to the specific selection criteria adopted in that study (see discussion above).
 
At $\feh > -1.5 \dex$, Thamnos and GSE appear to follow distinct trends, whereas at lower metallicity their distributions increasingly overlap. As is observed also in the light elements (see Fig. \ref{fig:abu_light}), Thamnos exhibits enhanced abundances in all $\alpha$-elements (Mg, Ca, and Ti) compared to GSE by up to $0.2 \dex$ at $\feh > -1.5 \dex$. Indeed, the $\alpha$-element abundances in Thamnos are consistent with those observed in the MW high-$\alpha$ sequence \citep{nissen&schuster2010}, showing a flat trend up to $\feh \sim -0.5 \dex$ and hinting at a strong in situ contamination in the higher-metallicity regime. On the contrary, GSE \afeh ratios start to drop at $\feh \sim -1.2 \dex$, reflecting a non-negligible contribution by SN Ia to the chemical enrichment of the gas already in place at this metallicity, likely due to a less efficient star formation compared to the MW \citep{vincenzo2019,gallart19,gonzalez-koda2025}.

\subsection{Abundances of iron-peak elements}

The synthesis of iron-peak elements arises from multiple nucleosynthetic pathways, with CCSNe, hypernovae (HNe)\footnote{These events represent a class of massive ($\geq$ 20 $\rm M_{\odot}$) CCSNe with $10$ times or more higher explosion energies relative to standard ones ($E=10^{51}$ erg, e.g. \citealt{umeda2002,Koba06}).} and SN Ia each contributing to a different extent to their overall production \citep{romano2010,kobayashi2020}. Results for the iron-peak elements are shown in Fig. \ref{fig:abu_ironpeak}. The chemical trends of GSE match those reported in \citetalias{ceccarelli2024}, and the relative abundances to Thamnos resemble the differences observed between low- and high-$\alpha$ population in the MW \citep{nissen&schuster2010,nissen&schuster2011}. In particular, Thamnos is on average enhanced in Cr, Co, Ni, and Zn compared to GSE at fixed metallicity, especially at $\feh > -1.5 \dex$. Among the elements we analysed, the largest difference ($\sim 0.3 \dex$) arise in the weak $s$- process Cu. Indeed, [Cu/Fe] abundances are significantly sub-solar ($\le -0.5$ dex) in both structures at [Fe/H] $\sim -2$ dex, but increase with metallicity, with a steeper rise in Thamnos candidate stars compared to GSE. From a theoretical perspective, Cu is mainly produced in massive stars via the weak $s$- process, and its yield is expected to grow with metallicity due to the higher availability of metal seed nuclei \citep{romano&matteucci2007,prantzos2018}. The shallower [Cu/Fe] increase in GSE suggests a reduced contribution from massive stars, consistent with theoretical predictions and observations in other dwarf galaxies, such as the Magellanic Clouds \citep{vanderswaelmen2013,mucciarelli2023}. However, we caution that NLTE effects can be significant, especially at low metallicity, limiting the reliability of Cu as a precise chemical tagger \citep[e.g.][]{andrievsky2018}. We see no difference in the trends of Thamnos and GSE in Mn. All of this evidence combined suggests that the chemical enrichment of the interstellar medium where dynamically selected Thamnos stars with $\feh > -1.5 \dex$ formed, was more strongly influenced by massive stars or did not receive significant contribution by SN Ia. This in turn highlights an origin from a stronger star-forming environment than that of GSE.

\subsection{Abundances of neutron capture elements}

This family of elements is divided into two classes depending on the efficiency of neutron capture compared to the timescale of the $\beta$ decay. In particular, slow neutron capture ($s$-process) elements are predominantly synthesised in either massive stars and intermediate-mass AGB stars (light and/or weak $s$-process elements, e.g. Y) or low mass AGBs (heavy and/or main $s$-process elements, e.g. Ba and La) AGB stars \citep{kobayashi2020}. In contrast, rapid ($r$-) neutron capture elements (e.g. Eu) are produced through a variety of astrophysical events, including magneto-rotational supernovae \citep{mosta2018} and/or collapsars \citep{siegel2019}, and neutron star mergers \citep[NSMs,][]{lattimer1974,argast04}.

Fig. \ref{fig:abu_neutroncapture} presents the abundance trends for neutron capture elements. We observe a significantly larger scatter in the [YII/Fe], [LaII/Fe], and [BaII/Fe] ratios for both Thamnos and GSE, relative to the tighter trends seen in the $\alpha$- and iron-peak elements (see Figs. \ref{fig:abu_light} $-$ \ref{fig:abu_ironpeak}), consistent with previous findings by \citet{matsuno22}. This increased dispersion is likely attributable to the combined contributions from the light, heavy $s$- and $r$- process formation channels involved in the synthesis of these elements \citep{kasen2017}. We identify an excess of stars enhanced in all $s$-process elements (i.e. [YII/Fe] $> 0.2$ dex, [LaII/Fe] $> 0.5$ dex, and [BaII/Fe] $> 0.5$ dex) within the metallicity range $-2.0 < \mathrm{[Fe/H]} < -1.0$ dex, which may indicate a binary origin. Nevertheless, all of these stars exhibit reliable astrometric solutions from \textit{Gaia} \citep[\texttt{RUWE} $< 1.4$,][and references therein]{el-badry2024}, while only two have a spectroscopic $V_{\mathrm{los}}$ that differs from the \textit{Gaia} values at the $3\sigma$ level. To probe the relative efficiency of these processes, we examined the [YII/BaII] and the [EuII/BaII] ratios (see top and middle panels of Fig. \ref{fig:abu_neutroncapture2}). The [YII/BaII] ratio shows, on average, lower values in GSE stars at $\feh > -1.5 \dex$, which have also been observed by various authors when comparing to in situ stars \citep{nissen&schuster2011,matsuno2021}. This trend may reflect a reduced efficiency in the production of light $s$-process elements in GSE, which is corroborated by the depletion observed in other weak $s$-process tracers such as Cu (see Fig. \ref{fig:abu_ironpeak}). Also, we can attribute the lower [YII/BaII] to more prominent contribution of strong $s$-process production by low-mass stars at equivalent metallicity, as the slower pace at which chemical enrichment proceeded in GSE allowed a more efficient enrichment by delayed stellar populations. Indeed, this rise is not observable in the [BaII/Fe] plot due to more prominent contribution by SN Ia, which have a similar timescale of enrichment to that of low-mass stars, preventing a rise in the [BaII/Fe]. Therefore, the lower light-to-heavy $s$ ratios at $\feh > -1.5 \dex$ can be explained by a combination of smaller production from the weak $s$- process and larger production in the strong $s$- process. Concerning the $r$-process contribution, the GSE population shows enhanced production through this channel, which is also mirrored by its enhancement in Eu, a pure tracer of the $r$- process (see Fig. \ref{fig:abu_neutroncapture}). Indeed, among the neutron capture species, Eu exhibits the clearest distinction between Thamnos and GSE: stars with $\feh > -1.5 \dex$ in the Thamnos substructure are on average depleted in [EuII/Fe] by $\sim 0.3 \dex$ compared to GSE candidates. This appears to be a common feature when comparing the abundance patterns of MW satellites with that of MW in situ stars \citep[e.g.][]{palla25}. Indeed, the Eu abundances in our sample, derived from the EuII lines at 4129 \r{A} and 6645 \r{A}, reach [EuII/Fe] $\gtrsim 0.4 \dex$ in GSE stars, consistent with prior determination in the literature for samples of either GSE stars \citep{aguado21,matsuno2021,naidu2022eu} and globular clusters \citep{ceccarelli2024b,monty2024}, and resembles the enhancement observed in surviving dwarf galaxies \citep{letarte2010,lemasle2014,liberatori25}. The different [EuII/Fe] patterns can explain the different behaviour in [EuII/BaII] shown by Thamnos candidates and GSE, with the latter enhanced relative to the former at [Fe/H] $> -1.5$ dex. This means that the enhancement in $r$- process (traced by Eu) is in proportion more efficient than the one from the strong $s$- process (traced by Ba). Moreover, the distinct trends exhibited by Thamnos and GSE stars in the [EuII/Mg] abundance plane at $\feh > -1.5 \dex$ (see bottom panel of Fig. \ref{fig:abu_neutroncapture2}) provide compelling evidence for the differing star formation timescales and efficiencies that characterised their respective progenitor environments at these high metallicities. In fact, despite Eu being a mixture of CCSNe-like and delayed events \citep[see][]{cote2019,molero2023}, recent evidence highlights that the main source of $r$- process at $\feh \lesssim -1 \dex$ are NSM \citep{palla25}. Therefore, given that in the observed regime Mg and Eu trace nucleosynthetic sources operating on markedly different timescales (i.e. CCSNe and delayed $r$-process events, respectively), the observed depletion further supports the conclusion that stars in the Thamnos substructure with $\feh > -1.5 \dex$ experienced a star formation history (SFH)more consistent with that of in situ populations. On the other hand, for $\feh < -2$ dex, the abundances are more consistent with those of GSE, with possibly slightly larger scatter (in e.g. [Mg/Fe], [TiII/Fe], [YII/BaII], and [EuII/Mg].) \\
  
\section{Comparison with galactic chemical evolution models}
\label{sec:GCE_models}
\begin{figure*}[!th]
\begin{minipage}{0.33\textwidth}
\centering
\includegraphics[width=1.0\textwidth]{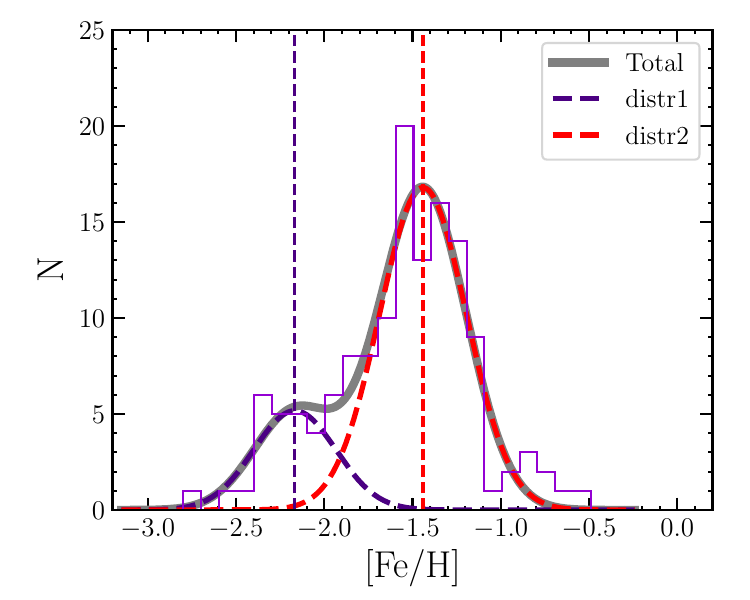} 
\end{minipage}
\begin{minipage}{0.33\textwidth}
\centering
\includegraphics[width=1.0\textwidth]{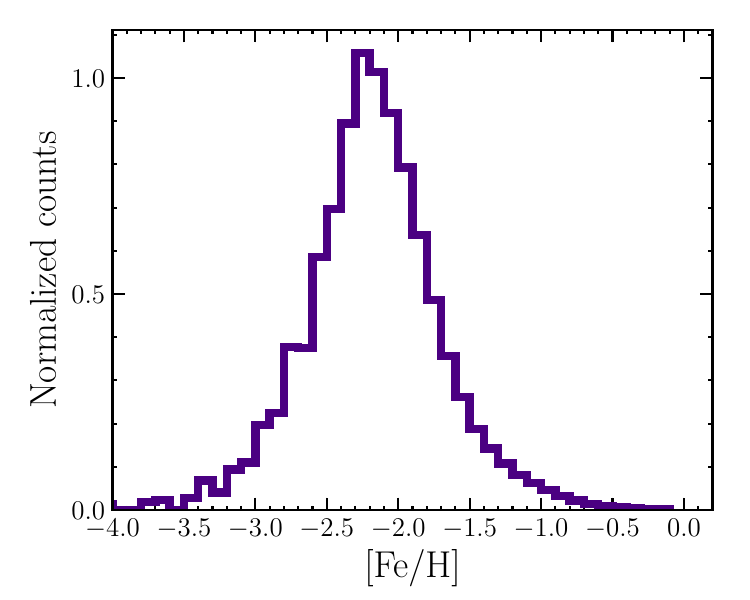}
\end{minipage}   
\begin{minipage}{0.33\textwidth}
\centering
\includegraphics[width=1.0\textwidth]{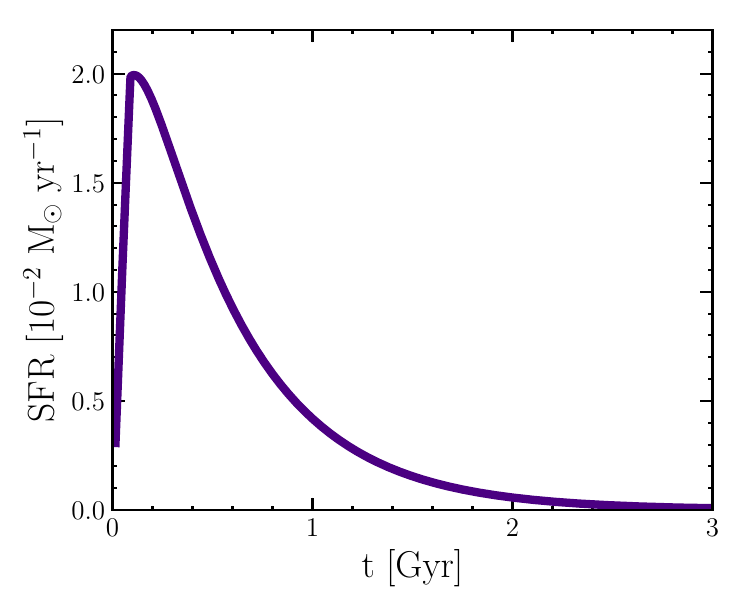}
\end{minipage}  
\caption{Division of the MDF of the Thamnos substructure (see Fig. \ref{fig:mdf} bottom panel) into two distributions to identify the $\lq$true' Thamnos component and the contamination from in situ stars (left panel) and MDF, SFH as predicted by the model built for Thamnos progenitor presented in Section \ref{subsec:chem_evo_assumptions} (two rightmost panels). In the left panel, the $\lq$true' Thamnos distribution traced by observation is labelled with $\lq$distr1' (dashed blue line), while MW in situ contamination would be represented by $\lq$distr2' (dashed red line).}
\label{fig:models_Thamnos}%
\end{figure*} 
\begin{figure*}[!th]
\centering
\begin{minipage}{0.45\textwidth}
\centering
\includegraphics[width=1.0\textwidth]{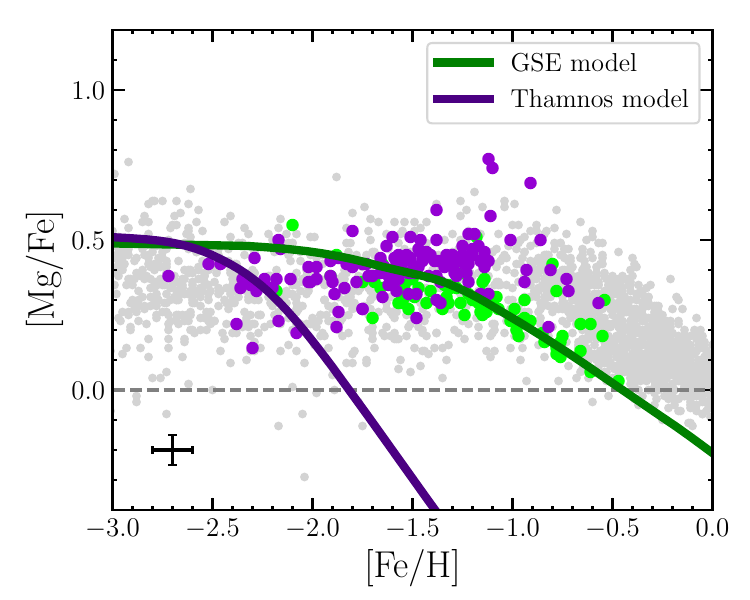} 
\end{minipage}
\begin{minipage}{0.45\textwidth}
\centering
\includegraphics[width=1.0\textwidth]{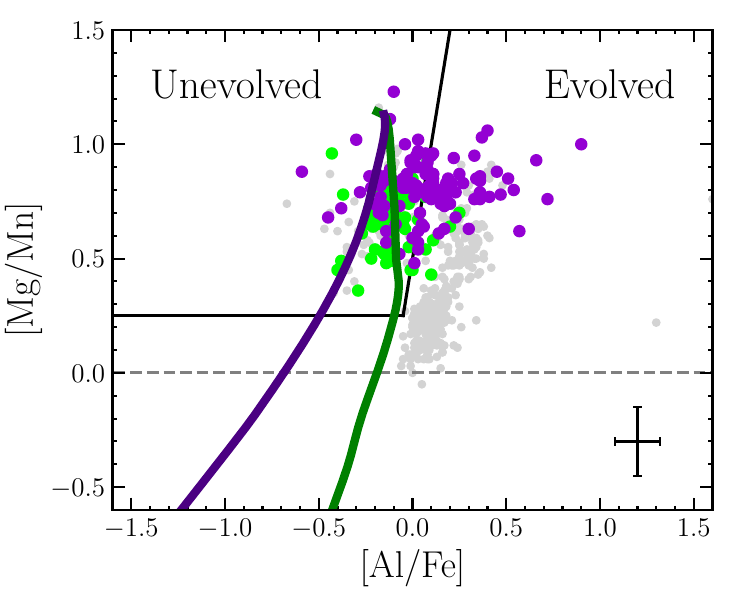}
\end{minipage}   
\caption{[Mg/Fe] - [Fe/H] (left panel) and [Mg/Mn] - [Al/Fe] (right panel) as predicted by the GCE models presented in Section \ref{sec:GCE_models} compared with the abundances measured in Thamnos and GSE candidates (purple and green points, respectively). Solid lines identify model tracks for GSE (green) and Thamnos (purple). In the right panel, the thin black line represents the division between chemically $\lq$unevolved' (top left) and $\lq$evolved' regions (top right) presented in \citet{horta2021}.}
\label{fig:models}%
\end{figure*} 

The chemical patterns observed in many chemical elements (e.g. $\alpha$-elements, Na, Al, Cu, Zn, and Eu) for the dynamically selected Thamnos stars with $\feh > -1.5 \dex$ are reminiscent of those typically measured for in situ stars, and stand in contrast to the abundances generally found in stars associated with accreted substructures or surviving dwarf galaxies at this metallicity \citep[][\citetalias{ceccarelli2024}]{venn2004,tolstoy09,nissen&schuster2010,aguado21,belokurov22,matsuno22,horta23}. In light of these results, it is likely that the Thamnos sample is significantly contaminated by in situ stars, especially in the higher-metallicity regime. To further investigate this hypothesis on a more quantitative basis, we developed GCE models to reproduce and interpret the chemical trends observed in these substructures.

In Appendix \ref{app:B} we briefly recap the main ingredients of GCE models. For more details on the general model structure and equations, we direct the reader to \citet{Palla20a,Matteucci21}. 

\subsection{Models for GSE and Thamnos progenitors}
\label{subsec:chem_evo_assumptions}

Below, we summarise the main model assumptions for the adopted GCE models tailored reproduce the available observables for the GSE and Thamnos progenitors. 
It is worth noting that, when possible, the adopted assumptions anchor on previously tested GCE modelling for the specific objects. 

\begin{itemize}
    \item GSE progenitor. The GCE model adopted for the GSE galaxy progenitor is analogous to the one proposed by \citet{vincenzo2019}.
    This model is characterised by a mild star formation efficiency ($\nu \simeq 0.4$ Gyr$^{-1}$, i.e. $\sim \times 5$ smaller than the one adopted for modelling the MW halo and thick disc, see e.g. \citealt{Spitoni19}), a very fast infall timescale ($\tau_{inf} \sim 0.25$ Gyr), and a mild outflow efficiency ($\omega=0.5$). For more details, we refer to the original paper of \citet{vincenzo2019}.

    \item Thamnos progenitor. We built a model for a Thamnos progenitor based on the information from the MDF provided in this work. 
    To avoid contamination from, likely, the Aurora population and catch the signature of the $\lq$true Thamnos' (see Section \ref{subsec:MDFs}), we fixed the GCE model parameters in order to reproduce the peak at $\feh < -2.0 \dex$ in the MDF of the Thamnos substructure. To isolate such a component, we assumed the MDF shown in Fig. \ref{fig:mdf} lower panel as composed by the sum of two Gaussian distributions and consider the low-metallicity component ($\lq$distr1' in the left panel of Fig. \ref{fig:models_Thamnos}) as the the putative $\lq$true Thamnos' MDF. 
\end{itemize}

We are able to reproduce $\lq$distr1' with a model with small gas accretion timescales ($\sim 0.5$ Gyr) and large wind mass loading factor ($\sim 10$), combined by a modest star formation efficiency ($\sim 0.1$ Gyr$^{-1}$). These ingredients lead to a very short star-forming episode (see right panel of Fig. \ref{fig:models_Thamnos}), with most of the stellar mass ($\sim 85$\%) built during the first gigayear of evolution. This is comparable to the estimate that the $\lq$true Thamnos' already built half of its mass $12.3 \pm 0.3$ Gyr ago \citep{dodd25SFH}. Such a SFH allows one to keep a MDF peaking at low metallicity ($\feh \sim -2.15 \dex$) with a relatively narrow distribution ($\pm 1 \sigma \sim 0.45 \dex$), as is shown in the central panel of Fig. \ref{fig:models_Thamnos}.

Fig. \ref{fig:models} shows the predicted abundance tracks by the models described above for GSE (in green curve) and for the $\lq$true Thamnos' (in purple) compared with the observations  for [Mg/Fe] - [Fe/H] (left panel) and [Mg/Mn] - [Al/Fe] (right panel).
These chemical diagnostic diagrams can be used as a powerful probe for the SFH of galaxies as shown by the time-delay model (\citealt{Matteucci12,hawkins2015,das2020,Matteucci21}). As is shown in the left panel of Fig. \ref{fig:models}, the predicted chemical track for the GSE progenitor aligns well with the observed trend, following tightly the [Mg/Fe] evolution up to the metal-rich end of GSE stars. The latter fall below the general trend for MW stars (grey points), highlighting an environment with less efficient star formation. 
On the other hand, the predicted [Mg/Fe] evolution for the Thamnos progenitor (purple line) allows us to identify the trend for putative $\lq$true Thamnos' stars. Indeed, the model closely follows the metal-poor end of the observed stars, whose [Mg/Fe] level is depleted relative to the bulk of the dynamically selected Thamnos sample, which shows instead a [Mg/Fe] plateau similar to what seen in the MW in situ halo. A possible, qualitative indication emerging from the model is that Thamnos stars might exhibit nearly solar [Mg/Fe] at metallicities around [Fe/H] $\sim -2$ dex. However, only with larger and more statistically significant samples will it be possible to spectroscopically confirm this prediction.

The likely heavy contamination of MW in situ stars in the Thamnos substructure can also be noted in the right panel of Fig. \ref{fig:models}, where we plot the [Mg/Mn] - [Al/Fe] space. This diagnostic diagram is divided into regions corresponding to chemically distinct stellar populations, as is defined by \citep{horta2021}. In particular, chemically $\lq$unevolved' stars are those that formed in environments that did not experience substantial chemical enrichment, and thus they are typically interpreted as either stars accreted from low-mass dwarf galaxies or those born in situ in the early proto-Galaxy. Conversely, the chemically $\lq$evolved' population shows abundances typical of stars formed in an environment that underwent efficient star formation, indicative of in situ formation \citep{horta25}. Indeed, a large fraction ($\sim 50\%$) of dynamically selected Thamnos stars fall in the $\lq$evolved' region, whereas model predictions (and the rest of the Thamnos stars) remain in the region of chemically $\lq$unevolved' stars, in agreement with the low SFE and therefore SFRs as for the Thamnos model. Less contamination from in situ stars is instead seen for GSE stars (only $\sim 10 \%$ of stars in the $\lq$evolved' region), with observations closely following the model predictions. 
The GSE model favours a system with stronger star formation than the putative $\lq$true Thamnos' progenitor, with [Mg/Fe] knee at larger metallicity and greater [Al/Fe] ratios, but still lower than the MW in situ population. 

\section{Discussion and conclusion}
\label{sec:discussion}

In this work, we have been able to fully characterise for the first time the chemical composition of a large sample of stars dynamically selected to be tentative members of Thamnos using high-resolution spectroscopy, comparing its abundance patterns with those of GSE and evaluating the observed trends against predictions from GCE models. We find that the dynamical region identified to have stars from Thamnos \citep[see][]{dodd23} has an overlap of populations with different chemistry. Indeed, on average, this dynamically identified Thamnos substructure is more metal-poor than GSE, yet exhibits systematically enhancement in the \afeh ratios in the more-metal rich regime ($\feh > -1.7 \dex$). This peculiar behaviour in the $\alpha$-elements has been previously tentatively noted based on limited samples from spectroscopic surveys with lower spectral resolution, such as APOGEE and LAMOST \citep{koppelman19,dodd23,horta23}. As this work represents the first dedicated high-resolution spectroscopic study of a large sample of stars in the Thamnos substructure, we firmly established the peculiarity of their chemical composition relative to other retrograde substructures. In particular, such enrichment in the \afeh ratios (see Fig. \ref{fig:abu_alpha}) is atypical for stars with similar \feh formed in low-mass dwarf galaxies, as is shown by the $\alpha$-element depletion commonly observed in remnants of minor merger, such as Sequoia or the Helmi streams \citep[][\citetalias{ceccarelli2024}]{matsuno22_helmi,matsuno22,horta23}, and in surviving dwarf galaxies \citep[see e.g.][]{tolstoy09,venn2012,hill2019}. As is shown in Fig. \ref{fig:abu_alpha}, a large fraction  of the stars in the Thamnos substructure lies along the high-$\alpha$ sequence, suggesting that these have in situ origin. This interpretation is further supported by the observed sharp rise in [Al/Fe] between $-1.7 < \feh < -1.0 \dex$ (see Fig. \ref{fig:abu_light}), a behaviour expected and observed in massive galaxies such as the MW \citep{hawkins2015,das2020,horta2021,belokurov22}. Additional support for in situ contamination comes from the enhancement of other elements produced by massive stars, such as Na, Co, Cu, and Zn. This is particularly apparent at $\feh > -1.5 \dex$, where stars in the Thamnos substructure overlap in metallicity with GSE, yet consistently follow in situ chemical trends. In contrast, GSE stars are either at the lower end of these distributions (e.g. Cr, Mn, Y) or appear to be depleted (e.g. Co, Ni, Cu, Zn). In the end, a notable case is that of Eu, a pure tracer of $r$-process nucleosynthesis. We find that stars dynamically selected to be in the Thamnos substructure are, on average, depleted in [EuII/Fe] by $\sim 0.3\dex$ compared to GSE at [Fe/H]$> -1.5$ dex. If this clump is indeed contaminated by in situ stars, then this trend is consistent with expectations: the high star formation efficiency in a massive galaxy such as the MW reduces the impact in Eu production by delayed $r$-process sources \citep[NSMs, e.g.][]{cescutti2015,matsuno2021,ou2024}, lowering the final [EuII/Fe] and [EuII/Mg] ratios, as is observed in Thamnos (see Fig. \ref{fig:abu_neutroncapture2}).

The metallicity distribution of stars in our dynamically selected Thamnos sample reveals a small bump at metallicities of $\feh < -2.0$ dex (see Fig. \ref{fig:mdf}), which we interpret as the potential signature of the accreted Thamnos dwarf galaxy. This is supported by the predictions of a GCE model tailored to fit this peak, which overlap with the observed abundances at these low metallicities (see Fig. \ref{fig:models}), for a system with a very low star formation efficiency and large wind mass loading, and that built most of its mass in the first gigayear of evolution, as is expected for an accreted dwarf galaxy. This model also highlights how such a system could not have the anomalously high \afeh values observed (see Fig. \ref{fig:models}), and supports the evidence of a significant amount of contamination in the Thamnos overdensity.

We estimated the level of contamination from in situ stars in the dynamically selected Thamnos sample using the MDF, as the ratio between the area under the more metal-rich Gaussian component ($\lq$distr2') and the total area under the best-fit curve, obtained by summing the two Gaussian components (see Fig. \ref{fig:models_Thamnos}). In this way we infer a contamination fraction of $\sim78\%$ across the entire metallicity range. This value is slightly higher than that estimated by \citet{dodd25SFH}, who reconstructed the SFH of Thamnos using CMD fitting. By comparing their results with mock samples with different levels of contamination by in situ stars on similarly bound orbits, they find an excess of metal-poor stars at $\feh \sim -2.0 \dex$ and infer a contamination lower than $50\%$, arguing that higher values would result in a too young and metal-rich population compared to the observed one. It is worth noting, however, that the contamination inferred in our work is not necessarily attributable exclusively to in situ stars, as a there is room for a contribution from GSE \citep[see also results by][based on APOGEE data]{mori2025}. This interpretation is in line with predictions from cosmological hydrodynamical simulations. Specifically, \citet{thomas25} report that contamination by in situ stars in substructures on bound and retrograde orbits such as Thamnos can range from $\sim 30\%$ up to $\sim 80\%$ for MW analogues in the Auriga suite of cosmological simulations \citep{grand2017}. Furthermore, they find that most of these substructures comprise stars accreted from two progenitor galaxies, with a typical population ratio of $\sim 30\%$, although this assertion might be a reflection of the limited resolution of their simulations, as they are unable to resolve dwarf galaxies with a stellar mass of  $\le 10^8 M_\odot$ (most being above $5 \times 10^8 M_\odot$). If such a scenario also applies to the MW, GSE would be the primary accreted contaminant within the Thamnos region of the IoM space. This is due to either the large range in energy and $L_{\mathrm{z}}$ covered by debris of such a massive dwarf galaxy \citep{koppelman2020,amarante2022,mori2024} and the presence of a rotating bar in the MW, which might push the low-energy tail of the GSE to more bound and more retrograde orbits, close to Thamnos \citep[][De Leo et al. in prep]{dillamore2025,woudenberg2025}. The possibility that Thamnos consists of two overlapping substructures (i.e. Thamnos 1 and 2) has been theorised since its discovery \citep{koppelman19,lovdal22,ruiz-lara22,bellazz23}. The two distinct peaks in the MDFs (see Fig. \ref{fig:mdf}) may offer observational evidence that these components trace two separate populations, with Thamnos 1 representing the more metal-poor, accreted dwarf galaxy and Thamnos 2 reflecting a population largely composed of contaminants. 
 
In summary, the contamination fraction inferred through our chemical analysis is in excellent agreement with previous estimates derived from fully independent methodologies, including dynamics and photometry. This concordance reinforces the idea that the majority of stars currently associated with the Thamnos substructure are likely of in situ origin or GSE contamination. Nevertheless, this work also suggests that the imprint of the original accreted dwarf galaxy may remain detectable within the metal-poor tail of the distribution, preserving valuable information on the early accretion history of the MW.

In the end, we note that GSE stars selected based on their dynamical properties exhibit coherent and well-defined chemical abundance trends across all the analysed chemical planes (see Figs. \ref{fig:abu_light} - \ref{fig:abu_neutroncapture}). As is shown in Fig. \ref{fig:models}, the derived abundances well match predictions from a GCE model for a GSE-like galaxy \citep{vincenzo2019}. Thus, this work remarks once again the effectiveness of high-resolution spectroscopy in disentangling chemical patterns of different MW halo substructures and characterising their progenitor galaxies. Also, this result highlights that the dynamical classification provided by \citet{dodd23} yields a remarkably pure GSE sample, with minimal contamination from non-accreted components. Indeed, within our sample, only nine stars at $\feh \ge -0.8 \dex$ deviate significantly from the mean GSE distributions, showing enhancement in several elements (Na, Al, Mg, Ti, Mn, Co, Ni, Cu, and Zn) and falling in the $\lq$evolved' regime in the [Mg/Mn]-[Al/Fe] space (see right panel of Fig. \ref{fig:models}). The abundances derived for these few stars align with the chemical trends observed in the metal-rich regime of Thamnos, and therefore we interpret them as likely contaminants from the in situ stellar population. 

To conclude, the results of this study emphasise once again the complex and heterogeneous nature of stellar substructures in the MW halo and the essential role of chemical abundances in disentangling their origins. While dynamics offers a strong starting point, it is only by adding independent analysis, such as high-resolution spectroscopy, that the nature of these systems can be robustly assessed. The forthcoming \textit{Gaia} DR4, combined with next-generation spectroscopic surveys such as 4MOST \citep{deJong2019} and WEAVE \citep{dalton2020}, will be pivotal in extending such analyses to more distant and fainter stars. These efforts will dramatically improve our ability to resolve the evolutionary history of the MW by enabling a more complete and detailed chemodynamical mapping of its accreted and in situ components. 

\section*{Data availability}

Tables \ref{tab:op}, \ref{tab:parameters} and \ref{tab:abundances} are only available in electronic form at the CDS via anonymous ftp to cdsarc.u-strasbg.fr (130.79.128.5) or via \url{http://cdsweb.u-strasbg.fr/cgi-bin/qcat?J/A+A/}.

\begin{acknowledgements}

Based on observations collected at the ESO-VLT under the programs 0112.B-0236 (P.I. E. Ceccarelli) and 0113.B.0196 (P.I. E. Ceccarelli).

MB, EC, and DM acknowledge the support to this study by the PRIN INAF 2023 grant ObFu \textit{CHAM - Chemodynamics of the Accreted Halo of the Milky Way} (P.I. M. Bellazzini).
    
This research is funded by the project \textit{LEGO – Reconstructing the building blocks of the Galaxy by chemical tagging} (P.I. A. Mucciarelli), granted by the Italian MUR through contract PRIN 2022LLP8TK\_001.

DM acknowledges financial support from PRIN-MIUR-22 ``CHRONOS: adjusting the clock(s) to unveil the CHRONO-chemo-dynamical Structure of the Galaxy” (PI: S. Cassisi) granted by the European Union - Next Generation EU.    

MB and DM acknowledge the support to activities related to the ESA/\textit{Gaia} mission by the Italian Space Agency (ASI) through contract 2018-24-HH.0 and its addendum 2018-24-HH.1-2022 to the National Institute for Astrophysics (INAF). 

AH acknowledges financial support through a Spinoza Award from  NWO (SPI 78-411). 

This work has made use of data from the European Space Agency (ESA) mission \textit{Gaia} \url{https://www.cosmos.esa.int/gaia}), processed by the \textit{Gaia} Data Processing and Analysis Consortium (DPAC, \url{https://www.cosmos.esa.int/web/gaia/dpac/consortium}). Funding for the DPAC has been provided by national institutions, in particular the institutions participating in the \textit{Gaia} Multilateral Agreement. 

This research has used data, tools or materials developed as part of the EXPLORE project that has received funding from the European Union’s Horizon 2020 research and innovation programme under grant agreement N° 101004214.

\end{acknowledgements}

\bibliographystyle{aa}
\bibliography{WRSII_thamnos.bib}

\begin{appendix}
\section{Methods}\label{app:A}

In the following we briefly describe the methods we use to derive orbital parameters and chemical abundances for the entire sample. A complete description of the procedure can be found in Sections 3 and 5 of \citetalias{ceccarelli2024}, respectively.

\subsection{Dynamics of the sample}\label{app:op}
We combined the 5D phase space information from \textit{Gaia} DR3 with the line-of-sight velocity ($V_{\mathrm{los}}$) obtained from high-resolution spectra via cross-correlation with template spectra with the same atmospheric parameters (i.e. \Teff, log $g$, \vt, [Fe/H]) through \texttt{fxcor} from \texttt{IRAF}. The uncertainties on the final $V_{\mathrm{los}}$ are computed as discussed in \citet{tonry&davies1979}, and they are typically $< 0.4$ km s$^{-1}$. We corrected both parallax and $V_{\mathrm{los}}$ for the effects of the \textit{Gaia} zero-point offset \citep{lindegren21} and the gravitational redshift \citep{zwitter18}, respectively. Within the sample, we find 9 stars with a difference with the \textit{Gaia} $V_{\mathrm{los}}$ measurement larger than 3$\sigma$, and we flag them as potential binaries in the final catalogue.

Stellar orbits were reconstructed using the software \texttt{AGAMA} \citep{vasiliev19} assuming the MW potential described in \citet{mcmillan17} and the same reference frame as in \citetalias{ceccarelli2024}. The final values of the orbital parameters are computed as the median of $100$ Monte Carlo realisations of each orbit assuming Gaussian distributions for the uncertainties in proper motions, parallax and $V_{\mathrm{los}}$, with associated uncertainties at the $16$th and $84$th percentiles. Final values are listed in Table \ref{tab:op}.

\subsection{Atmospheric parameters}\label{app:ap}
First input effective temperatures (\Teff) have been estimated following the colour - temperature relation by \citet{mucciarelli21} using the $(BP - RP)$ colour, assuming $E(B-V)$ from the \texttt{EXPLORE} tool and $\feh = -1.5 \dex$. Surface gravities (log $g$) have been derived through the Stefan-Boltzmann relation and microturbulent velocities (\vt) have been obtained by imposing no the trend between iron abundances and reduced equivalent widths. As the colour-temperature relation depends on metallicity, first estimates for the stellar parameters were subsequently refined adopting the correct \feh values. Derived atmospheric parameters for target stars are listed in Table \ref{tab:parameters}.

\subsection{Chemical analysis}\label{app:ca}
Abundances are derived using \texttt{ATLAS9} \citep{kurucz} model atmospheres and the linelists with atomic and molecular data from the Kurucz/Castelli\footnote{\url{http://wwwuser.oats.inaf.it/castelli/linelists.html}} database. The abundance analysis for Na, Mg, Ca, Ti, Cr, Fe, Ni, and Zn is based on the equivalent width (EW) method, using the code \texttt{GALA} \citep{gala}. EWs are measured with \texttt{DAOSPEC} \citep{daospec}. For the elements with hyperfine/isotopic splitting transitions and/or blended lines (Al, Mn, Co, Cu, Y, Ba, La and Eu), the abundances were derived via spectral synthesis using the code \texttt{SALVADOR} (Alvarez Garay et al. in prep). All the abundance ratios refer to the solar composition by \citet{grevesse1998}. The procedure for error estimates is described in detail in Section 5.4 of \citetalias{ceccarelli2024}.

\section{Basic ingredients for GCE models}\label{app:B}

As commonly assumed in the GCE literature \citep{Romano13,vincenzo2019,Koba20SNIa}, to fuel star formation cold gas of primordial chemical composition is accreted to the galaxy at an exponentially decreasing rate:
\begin{equation}
    \Dot{M}_{inf}(t) \propto e^{-t/\tau_{inf}},
    \label{eq:gas_infall}
\end{equation}
where $M_{inf}(t)$ is the mass accreted at time $t$ and $\tau_{inf}$ the e-folding time for gas accretion. 
The star formation rate (SFR) is implemented according to the Kennicutt-Schmidt law \citep{Kennicutt98}: 
\begin{equation}
    \psi(t)= \nu M_{gas}(t)^k,
    \label{eq:SFR}    
\end{equation}
with $k=1$ and the star formation efficiency (SFE) $\nu$ as the control parameter that represents the SFR per unit mass of gas, $M_{gas}(t)$. 

As we are dealing with systems with shallower potential wells due to their fairly low masses, the models also allow for gas loss through outflows. Here, we assume them to be proportional to the SFR (e.g. \citealt{Vincenzo15,Molero21,Palla24b}):
\begin{equation}
    \Dot{M}_{out}(t) = \omega \, \psi(t),
    \label{eq:outflow}
\end{equation}
where $\omega$ is the mass loading factor parameter.

For what concerns the stellar nucleosynthesis, all the models adopted in this work relax the instantaneous recycling approximation, therefore allowing the different elements to be restored to the interstellar medium according to the lifetimes of their stellar progenitors. These are weighted according to the stellar initial mass function, for which we use the one by \citet{Kroupa93}, extensively used to model the evolution of the MW components and its satellites (e.g. \citealt{Romano05,vincenzo2019,Palla20,Nieuwmunster23}).

We adopt well-tested yield sets by \citet{Karakas10} for low- and intermediate-mass stars, \citet{nomoto2013} for massive stars/CCSNe and \citet{Iwa99} for SN Ia. For the latter, the delay-time-distribution by \citet[see also \citealt{Palla21} for details]{MatteucciRecchi01} is assumed.
For \citet{nomoto2013} massive star yields instead, we assume a decreasing HN fraction with metallicity as in \citet{mucciarelli2021NatAs}.

\end{appendix}

\end{document}